\documentclass[twocolumn,tight,times]{aastex631}

\usepackage{graphicx}	
\usepackage{color,soul}

\usepackage{amsmath}	
\usepackage{amssymb}	
\usepackage{amsbsy}	
\usepackage{newtxtt,newtxmath}
\usepackage[T1]{fontenc}
\usepackage{catchfile}
\usepackage{enumitem}
\usepackage{longtable}

\usepackage{xcolor}
\usepackage{mathtools}
\usepackage{hyperref}
\usepackage{array}

\newcommand{\cgsline}{erg s$^{-1}$ cm$^{-2}$}

\newcommand{\msol}{M$_\odot$}

\newcommand\C[1]\null

\newcommand{\halpha}{H$\alpha$}
\newcommand{\hbeta}{H$\beta$}

\newcommand{\nii}{[N{\sc ii}]}
\newcommand{\sii}{[S{\sc ii}]}
\newcommand{\oii}{[O{\sc ii}]}
\newcommand{\oiii}{[O{\sc iii}]}

\newcommand{\logoh}{$12+\log(\mathrm{O/H})$}

\begin{document}

\title{The AURORA Survey: The Mass -- Metallicity and Fundamental Metallicity Relations at $z \sim 2.3$ \\ Based Purely on Direct $T_e$ Metallicities}

\correspondingauthor{Ali Ahmad Khostovan}
\email{akhostov@gmail.com}

\author[0000-0002-0101-336X]{Ali Ahmad Khostovan}
\affiliation{Department of Physics and Astronomy, University of Kentucky, 505 Rose Street, Lexington, KY 40506, USA}

\author[0000-0003-4792-9119]{Ryan L. Sanders}
\affiliation{Department of Physics and Astronomy, University of Kentucky, 505 Rose Street, Lexington, KY 40506, USA}

\author[0000-0003-3509-4855]{Alice E. Shapley}
\affiliation{Department of Physics \& Astronomy, University of California, Los Angeles, 430 Portola Plaza, Los Angeles, CA 90095, USA}

\author[0000-0001-8426-1141]{Michael W. Topping}
\affiliation{Steward Observatory, University of Arizona, 933 N Cherry Avenue, Tucson, AZ 85721, USA}

\author[0000-0001-9687-4973]{Naveen A. Reddy}
\affiliation{Department of Physics \& Astronomy, University of California, Riverside, 900 University Avenue, Riverside, CA 92521, USA}

\author[0000-0002-8111-9884]{Alex M. Garcia}
\affiliation{Department of Astronomy, University of Virginia, 530 McCormick Road, Charlottesville, VA 22904}
\affiliation{Virginia Institute for Theoretical Astronomy, University of Virginia, Charlottesville, VA 22904, USA}
\affiliation{The NSF-Simons AI Institute for Cosmic Origins, USA}

\author[0000-0002-4153-053X]{Danielle A. Berg}
\affiliation{Department of Astronomy, The University of Texas at Austin, 2515 Speedway, Stop C1400, Austin, TX 78712, USA}
\affiliation{Cosmic Frontier Center, The University of Texas at Austin, Austin, TX 78712, USA}

\author[0000-0003-1249-6392]{Leonardo Clarke}
\affiliation{Department of Physics \& Astronomy, University of California, Los Angeles, 430 Portola Plaza, Los Angeles, CA 90095, USA}

\author[0000-0002-3736-476X]{Fergus Cullen}
\affiliation{Institute for Astronomy, University of Edinburgh, Royal Observatory, Edinburgh, EH9 3HJ, UK}

\author[0000-0001-7782-7071]{Richard S. Ellis}
\affiliation{Department of Physics \& Astronomy, University College London, Gower Street, London WC1E 6BT, UK}

\author[0000-0003-4264-3381]{N. M. F\"orster Schreiber}
\affiliation{Max-Planck-Institut f\"ur extraterrestrische Physik (MPE), Giessenbachstr.1, D-85748 Garching, Germany}

\author[0000-0002-3254-9044]{Karl Glazebrook}
\affiliation{Centre for Astrophysics and Supercomputing, Swinburne University of Technology, P.O. Box 218, Hawthorn, VIC 3122, Australia}

\author[0000-0001-5860-3419]{Tucker Jones}
\affiliation{Department of Physics and Astronomy, University of California Davis, 1 Shields Avenue, Davis, CA 95616, USA}

\author[0000-0003-4368-3326]{Derek J. McLeod}
\affiliation{Institute for Astronomy, University of Edinburgh, Royal Observatory, Edinburgh, EH9 3HJ, UK}

\author[0000-0003-4464-4505]{Anthony J. Pahl}
\affiliation{The Observatories of the Carnegie Institution for Science, 813 Santa Barbara Street, Pasadena, CA 91101, USA}

\author[0000-0002-5139-4359]{Max Pettini}
\affiliation{Institute of Astronomy, Madingley Road, Cambridge CB3 OHA, UK}

\author[0000-0002-5653-0786]{Paul Torrey}
\affiliation{Department of Astronomy, University of Virginia, 530 McCormick Road, Charlottesville, VA 22904}
\affiliation{Virginia Institute for Theoretical Astronomy, University of Virginia, Charlottesville, VA 22904, USA}
\affiliation{The NSF-Simons AI Institute for Cosmic Origins, USA}


\begin{abstract}

We present new constraints on the Mass -- Metallicity (MZR) and Fundamental Metallicity Relations (FMR) using a sample of 34 galaxies at $1.38\leq~z\leq~3.5$ (median $z=2.28$). These galaxies have direct $T_e$ measurements from \oiii4363\AA~and/or \oii7320,7331\AA~auroral emission lines detected with \textit{JWST}/NIRSpec as part of the AURORA survey. The detection of both oxygen auroral lines allows for dual-zone direct $T_e$ measurements and expands the dynamic range in \logoh~(7.68 to 8.65 dex), stellar mass ($10^{8}$ to $10^{10.4}$~\msol), and star-formation rate ($1$ to $100$ \msol~yr$^{-1}$) compared to previous direct $T_e$ studies of the high-redshift MZR and FMR. We characterize the $z\sim2$ MZR and find a slope of $0.27\pm0.04$ and normalization of $\textrm{\logoh}~=~8.44\pm0.04$ at $10^{10}$~\msol~with an intrinsic scatter of 0.10~dex, consistent with past strong-line MZR measurements. Comparisons with $z\sim2$ predictions from six simulations reveal that none reproduce our observed MZR normalization evolution between $z\sim0$ and $z\sim2$. This discrepancy suggests current models do not fully capture the chemical enrichment and feedback processes occurring at cosmic noon. However, all 34 galaxies are on or above the star-forming main sequence such that our sample may be biased towards lower \logoh~if the FMR persists at $z\sim2$. Correcting for this selection effect would increase O/H by $\approx0.1$~dex at 10$^{9.3}$ \msol~(the median mass of our sample) bringing our MZR into better agreement with that of \texttt{TNG}. Lastly, we find our $z\sim2.3$ sample is consistent with the $z\sim0$ FMR within 0.1~dex in O/H, indicating that the smooth secular mechanisms regulating chemical enrichment, star formation, stellar mass, and outflows were in place at cosmic noon.
\end{abstract}

\keywords{\href{http://astrothesaurus.org/uat/594}{Galaxy evolution (594)}, \href{https://astrothesaurus.org/uat/224)}{Chemical abundances (224)}, \href{https://astrothesaurus.org/uat/1031}{Metallicity (1031)}, \href{ttps://astrothesaurus.org/uat/734}{High-redshift galaxies (734)}, \href{https://astrothesaurus.org/uat/459}{Emission line galaxies (459)}}

\section{Introduction}

The metal content of the interstellar medium (ISM) provides a powerful tracer of the baryon cycle that regulates how galaxies grow over cosmic time. Gas-phase metallicities encode the cumulative effects of star formation, gas inflow, metal production, feedback-driven outflows, and recycling of enriched metals. Two key scaling relations between gas-phase metallicities (\logoh) and galaxy properties encapsulate these processes: the Mass -- Metallicity Relation (MZR; e.g., \citealt{Tremonti2004}) and the Fundamental Metallicity Relation (FMR; e.g., \citealt{Mannucci2010}), where the latter links star-formation rate (SFR), stellar mass, and gas-phase metallicity. Combined, these relations trace the underlying physical mechanisms that drive chemical enrichment, mix newly produced metals with accreted pristine gas, remove metals via outflows, and recycle enriched gas from the circumgalactic medium. Cosmic noon ($z \sim 1.5 - 3.5$) represents a key epoch where the Universe peaked in cosmic star-formation rate density and galaxies rapidly assembled their stellar mass (e.g., \citealt{Madau2014,Khostovan2015,Khostovan2016}). Measuring the MZR and FMR during this epoch provides crucial insight into how galaxies chemically matured during the most active cosmic phase of galaxy assembly and how these relations evolved across cosmic time.

The `gold standard' of metallicity measurements is the direct $T_e$ method, which requires the detection of auroral emission lines such as \oiii4363\AA~and \oii7320,7331\AA~\citep{Osterbrock2006}. The ratio of these transitions relative to a bright counterpart from the same ion (i.e., \oiii5007\AA, \oii3726,3729\AA) is sensitive to the electron temperature which is used to convert emission-line flux ratios into ionic abundances through well-known atomic physics. However, auroral lines are notoriously difficult to observe beyond the local Universe with only a handful of pre-\textit{JWST} studies constraining \logoh~using this approach (e.g., \citealt{Yuan2009,Christensen2012,Gburek2019,Gburek2023,Sanders2016_auroral,Sanders2020,Sanders2023}). A common alternative for inferring metallicities in star-forming galaxies at high redshift is to use strong-line calibrations that are based on either theoretical modeling (e.g., \citealt{Kewley2002,Kewley2019}) or empirical $T_e$-based data sets at low-$z$ (e.g., \citealt{Pettini2004,Marino2013,Curti2020}). 

Strong-line calibrations provide the added advantage of robust number statistics as they only require measurements of key emission lines that are easily detectable from ground-based observatories relative to auroral lines. These calibrations have enabled systematic investigations of both MZR and FMR at $z \sim 1.5 - 3.5$ (e.g., \citealt{Erb2006,Cullen2014,Maier2014,Zahid2014,Sanders2015,Henry2021,Sanders2021,Topping2021,Revalski2024,Jain2025,Stanton2025}). However, strong-line calibrations based on line ratios, such as N2 (\nii6583\AA/\halpha) and O3N2 ($\textrm{\oiii5007\AA}/\textrm{\hbeta} \times \textrm{\halpha}/\textrm{\nii6583\AA}$), can suffer from an intrinsic scatter of $\sim 0.1 - 0.2$ dex (e.g., \citealt{Pettini2004,Marino2013}) and have systematic offsets of up to $\sim 0.5 - 1$ dex in inferred metallicity depending on which calibration is applied (see \citealt{Kewley2019} for a review). This is further complicated by the fact that high-redshift galaxies differ in ISM conditions relative to $z \sim 0$ galaxies exhibiting harder ionizing spectra, higher ionization parameters, elevated electron densities, and more extreme excitation conditions (e.g., \citealt{Steidel2014,Hayashi2015,Sanders2016,Sanders2023b,Khostovan2016,Khostovan2021,Khostovan2024,Khostovan2025,Isobe2023,Topping2025}). Although past efforts have recalibrated strong-line relations using local analogs of cosmic noon galaxies (e.g., \citealt{Curti2017,Bian2018,Perez2021,Nakajima2022}), robust constrains on the MZR and FMR at high redshift require direct $T_e$ measurements spanning a broad range of \logoh, stellar mass, and SFR.

\textit{JWST} has provided the needed sensitivity and wavelength coverage to not only detect the sought after \oiii4363\AA~auroral line, but also several other auroral lines such as \oii7320,7331\AA, [S{\sc iii}]6314\AA, and \sii4070\AA~at $z>1.5$ (e.g., \citealt{Rhoads2023, Curti2023, Nakajima2023, Sanders2023,Sanders2024, Laseter2024,Morishita2024}). This enables multi-zone direct $T_e$ metallicity measurements of the different ionization phases in H{\sc ii} regions. It also allows for a sizable number of auroral detections to recalibrate strong-line diagnostics using representative samples of cosmic noon galaxies (e.g., \citealt{Nakajima2023,Laseter2024,Sanders2024,Sanders2025,Chakraborty2025,Scholte2025}). 

In this Letter, we present direct $T_e$ measurements of 34 cosmic noon galaxies at $1.38 \leq z \leq 3.50$ from the \textit{JWST}/NIRSpec AURORA survey \citep{Shapley2025}. Direct $T_e$ measurements are based on \oiii4363\AA~and/or \oii7320,7331\AA~auroral line detections, with 18/34 galaxies having both auroral lines detected. This represents one of the largest samples of direct $T_e$ measurements at cosmic noon. We use this sample to constrain the MZR and FMR at $z \sim 2.3$ and compare our measurements with previous strong-line calibration measurements and predictions from a suite of simulations. The paper is structured as follows: \S\ref{sec:data} describes the AURORA survey and ancillary multi-wavelength photometric data; \S\ref{sec:methodology} highlights our methodologies 
for deriving direct metallicities, stellar masses, and SFRs; \S\ref{sec:results} describes our results showing our MZR and FMR measurements; \S\ref{sec:discussion} discusses how our results compare to past observations and simulations and the physical implications; and, \S\ref{sec:conclusions} summarizes our main conclusions. 

Throughout this paper, we assume a \cite{Kroupa2001} initial mass function (IMF), $\Lambda$CDM ($\mathrm{H}_0 = 70$ km s$^{-1}$ Mpc$^{-1}$, $\Omega_m = 0.3$, $\Omega_\Lambda = 0.7$), and a solar oxygen abundance of 12+log(O/H$)_\odot=8.69$ \citep{Asplund2021}\footnote{This value is only used to express our derived O/H values in terms of a fraction of solar metallicity. Adopting the higher values of 12+log(O/H$)_\odot=8.74-8.76$ recently found by some groups \citep{Magg2022,Lodders2025} does not change our results.}.

\section{Data}
\label{sec:data}

\subsection{AURORA}
The Assembly of Ultradeep Rest-optical Observations Revealing Astrophysics (AURORA) survey (PID: 1914; PIs: A. Shapley \& R.~L.~Sanders) is a \textit{JWST}/NIRSpec program consisting of two pointings: GOODS-N and COSMOS. We refer the reader to the survey paper for details on observing strategy, target selection, data reduction, and spectral extraction \citep{Shapley2025}. Each pointing is observed with G140M/F100LP (44204s), G235M/F170LP (28886s), and G395M/F290LP (15056s) allowing for contiguous spectral coverage from $1 - 5$ \micron~(observer-frame) with $R \sim 1000$ and a sensitivity of $5 \times 10^{-19}$ \cgsline~(3$\sigma$). A total of 36 primary targets ($1.38 < z < 4.41$; $z_\textrm{med} = 2.33$) were selected based on ground-based spectroscopy of the strong rest-optical lines that suggested the auroral \oiii4363\AA~and/or \oii7320,7331\AA\ lines would be bright enough to detect in the AURORA integration times (12.3~h in G140M; 8.0~h in G235M; 4.2~h in G395M). Filler targets are also included and are described in detail in \cite{Shapley2025} with a total of 97 targets making up the full survey. Data reduction was done using the standard STScI \textit{JWST} pipeline coupled with custom software routines \citep{Shapley2025}. 1D science and error spectra were extracted using the optimal technique \citep{Horne1986}. Spectra were corrected for slit losses following \citet{Reddy2023,Reddy2025} and line fluxes were measured as described in \citet{Sanders2025}.

\subsection{Photometric Data}
\label{sec:photometry}
Ancillary photometric data the targets of this analysis were drawn from the DAWN \textit{JWST} Archive \citep[DJA;][]{Brammer2023,Valentino2023}. In COSMOS, this imaging covers $0.4$ -- $5$ \micron~in 15 filters: 7 \textit{HST} and 8 \textit{JWST}. \textit{JWST}/NIRCam coverage is drawn from PRIMER \citep{Donnan2024} and \textit{HST} imaging is from several programs including COSMOS/ACS \citep{Scoville2007}, CANDELS \citep{Grogin2011,Koekemoer2011, Nayyeri2017}, UVCANDELS \citep{Wang2025}, and 3D-\textit{HST} \citep{Skelton2014}. The imaging in GOODS-N consists of 9 \textit{HST} and 11 \textit{JWST} filters covering a wavelength range of $0.4$ -- $5$ \micron. The \textit{JWST} data is drawn from JADES \citep{Eisenstein2023}, CONGRESS (PID: 3577; PI: E. Egami), PANORAMIC \citep{Williams2025}, and FRESCO \citep{Oesch2023}. \textit{HST} imaging is a combination of several key programs including CANDELS \citep{Barro2019} and HDUV \citep{Oesch2018}.

\section{Methodology}
\label{sec:methodology}

\subsection{Sample selection}

We selected all star-forming galaxies in AURORA at $z\leq3.5$ with a signal-to-noise ratio ($\mathrm{S/N}$) $\ge3$ detection of at least one of the auroral lines \oiii4363\AA\ and \oii7320,7331\AA, enabling robust direct metallicities.
Quiescent galaxies and active galacitic nuclei (selected from either broad H$\alpha$ features in the absence of broad forbidden lines, or else $\log_{10} \textrm{\nii}/\textrm{\halpha}~>-0.3$) were excluded.
This selection resulted in a sample of 34 star-forming galaxies at $1.38 \leq z \leq 3.8$ (median $z\approx2.3$).
In total, there are 26 \oiii4363\AA~and 26 \oii7320,7331\AA~detections, including 8 galaxies with only \oiii4363\AA, 8 galaxies with only \oii7320,7331\AA, and 18 galaxies with detections of both oxygen auroral lines allowing for $T_e$ constraints of both the high- and low-ionization nebular zones. This sample covers $\sim 1$~dex in \logoh~ranging from 7.68 to 8.65 dex and $\sim2.5$~dex in stellar mass spanning $10^8-10^{10.5}$~$\mathrm{M}_\odot$.

\subsection{Direct $T_e$ Metallicity}

We refer the reader to \cite{Sanders2025} for details on how direct $T_e$ metallicities and associated errors were derived for the AURORA targets. In brief, electron density ($n_e$), electron temperature ($T_e$), and dust reddening ($E(B-V)_\mathrm{gas}$) were iteratively computed until convergence, using \texttt{PyNeb} \citep{Luridiana2015} for all atomic and ionic calculations. All Balmer and Paschen \ion{H}{1} lines detected at $\mathrm{S/N}\ge3$ were simultaneously fit to derive $E(B-V)_\mathrm{gas}$, assuming the \citet{Cardelli1989} extinction curve, and all line fluxes were subsequently corrected for dust attenuation. The \sii6716,6731\AA~doublet ratio was used to infer $n_e$ if both \sii\ lines were detected at $\mathrm{S/N}\ge3$. Otherwise, $n_e$ was estimated from the relation $n_e=40 (1+z)^{1.5}~\mathrm{cm}^{-3}$ \citep{Topping2025}.\footnote{The density inferred from \sii~probes the low ionization zone which is appropriate for $\mathrm{O}^{+}$, but may not reflect $n_e$ of the high $\mathrm{O}^{++}$ ionization zone. However, varying the assumed $n_e$ from 10 to 5000 cm$^{-3}$ negligibly changes the derived $T_e(\mathrm{O}^{++})$ by $<1$ percent.} The high ($T_e(\mathrm{O}^{++})$) and low ($T_e(\mathrm{O}^{+})$) ionization zone temperatures were computed from the \oiii4363\AA/5007\AA~and \oii7320,7331\AA/3726,3729\AA~line ratios, respectively. If either \oiii4363\AA~or \oii7320,7331\AA did not have $\mathrm{S/N}\ge3$, then $T_e$ for the corresponding ionization zone was estimated using the relation from \cite{Campbell1986}, found to be consistent with dual-zone $T_e$ constraints at $z\gtrsim2$ in AURORA \citep{Sanders2025} and other studies \citep{Cataldi2025,Chakraborty2025}. The ionic abundance ratios $\mathrm{O}^+/\mathrm{H}$ and $\mathrm{O}^{++}/\mathrm{H}$ were then inferred from the \oii3726,3729/H$\beta$ and \oiii5007/H$\beta$ ratios, respectively, using the $T_e$ and $n_e$ constraints. Total oxygen abundance was calculated as $\frac{\mathrm{O}}{\mathrm{H}} = \frac{\mathrm{O}^{+}}{\mathrm{H}} + \frac{\mathrm{O}^{++}}{\mathrm{H}}$.

\subsection{Stellar Mass}

The spectral energy distribution (SED) of each AURORA target was characterized using the multi-wavelength flux-calibrated imaging drawn from DJA (\S\ref{sec:photometry}). Images in each filter were point spread function (PSF) matched by convolving with a 2D Gaussian kernel with $\mathrm{FWHM}= \sqrt{\mathrm{FWHM}_\mathrm{filter}^2 - \mathrm{FWHM}_\mathrm{F160W}^2}$, where $\mathrm{FWHM}_\mathrm{F160W}=0.19''$ represents the largest PSF among the filter set. Flux densities were derived from forced photometry measurements using \texttt{photutils} \citep{photutils}, where Kron apertures were applied to the PSF-matched images in combination with the segmentation maps from DJA. We assumed a Kron scaling of 2.5 and minimum value for the unscaled Kron radius of 3.8 pixels. Photometric errors are based on the $1\sigma$ local background variation within the cutout, estimated by placing 1000 random apertures within the Kron area after masking out all sources using the segmentation maps.

We used the SED fitting code \texttt{Bagpipes} \citep{Carnall2018} to estimate stellar masses assuming \texttt{BPASS~v2.2.1} \citep{Stanway2018} binary stellar population models and a \cite{Kroupa2001} IMF with slopes $-1.35$ ($0.1 < M/\mathrm{M}_\odot < 0.5$) and $-2.35$ ($0.5 < M < 300$ $\mathrm{M}_\odot$). Stellar metallicities could vary from 0.001 -- 2 $Z_\odot$ but were limited by a Gaussian prior centered on \logoh~measured from AURORA with a width equivalent to the $1\sigma$ uncertainties on \logoh. We assumed a \citet{Calzetti2000} dust attenuation curve\footnote{We find that assuming an SMC dust attenuation curve instead decreases the resulting stellar masses by 0.08~dex on average, a negligible shift relative to the statistical uncertainties of our MZR measurements (see \S\ref{sec:MZR}).} with $0 < A_V < 5$~mag and a nebular-to-stellar reddening ratio between 1 and 3, with uniform priors on both parameters.

Nebular emission contributions within the medium and broadband photometry were accounted for via photoionization models incorporated in \texttt{Bagpipes}. We computed custom nebular grids using \texttt{Cloudy v17.03} \citep{Ferland2017} assuming \citet{Nicholls2017} abundance pattern scaling, ionization parameters of $-4 \leq \log_{10} U \leq -1$ with 0.2 dex increments, and $n_e = 10$, $50$, $100$, $500$, $1000$, and $1500$ cm$^{-3}$. For each target, the grid with density closest to the \sii-derived $n_e$ was adopted and the gas-phase metallicity was fixed to the value based on the derived direct $T_e$ metallicity. Lastly, we assumed a non-parametric star-formation history with the continuity prior \citep{Leja2019}. We adopted 7 lookback time bins bounded by 0, 3, 10, 30, 100, 300, 1000, and 3000~Myr that are of equal size in logarithmic space (0.5 dex), and a final bin extending to the age of the Universe at the spectroscopic redshift of the source.

\begin{figure}
    \centering
    \includegraphics[width=\columnwidth]{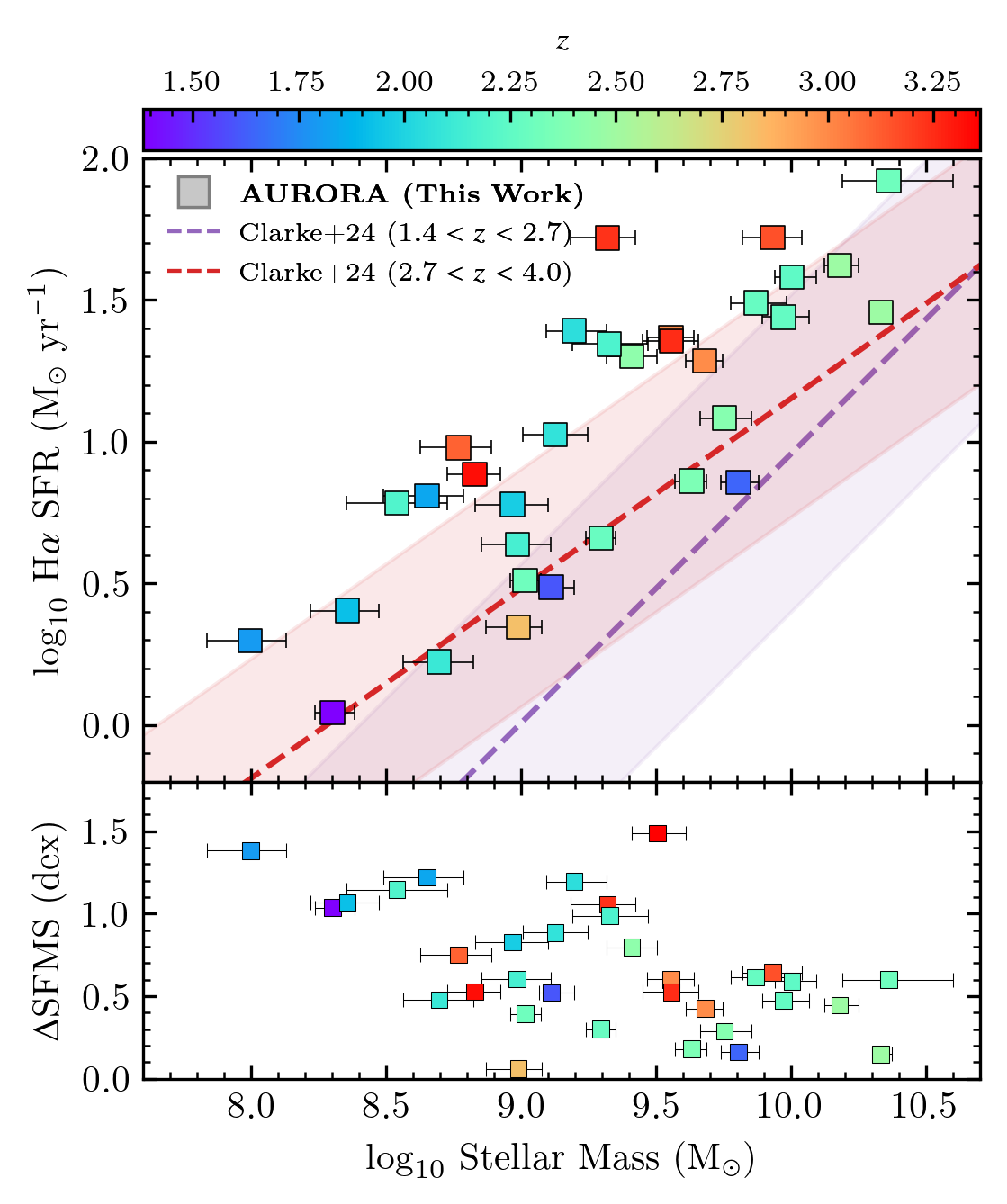}
    \caption{SFR vs. stellar mass for AURORA sources, shown in \textit{squares} color-coded by redshift. Overlaid are the \cite{Clarke2024} star-forming main sequence relations based on SFR(\halpha). Offsets from the \cite{Clarke2024} relations are shown in the \textit{bottom} panel. AURORA sources lie on and above the main sequence, reaching as high as $31$ times the typical SFR at fixed mass.}
    \label{fig:SFMS}
\end{figure}

\subsection{Star-Formation Rates}

SFRs were derived from the dust-corrected H$\alpha$ luminosity ($L_{\mathrm{H}\alpha}$) using a metallicity-dependent conversion factor \citep{Reddy2022,Reddy2023_SFR,Clarke2024} based on the relation presented in \citet{Sanders2025}:
\begin{equation}
     \frac{\mathrm{SFR}}{\textrm{\msol}~\textrm{yr}^{-1}} = \Big(\mathcal{C} \frac{L_{\mathrm{H}\alpha}}{\mathrm{erg~s}^{-1}}\Big) 10^{0.89\log_{10} Z + 0.14(\log_{10} Z)^2}
\end{equation}
where $Z=0.014\times 10^{12+\log(\mathrm{O/H}) - 8.69}$ is a function of the direct $T_e$ metallicity and $\mathcal{C} = 10^{-40.26}$. Figure \ref{fig:SFMS} shows the SFRs and stellar masses of our sample compared to the $1.4 < z < 4$ star-forming main sequence relation from \citet{Clarke2024}. All of our sources lie on or above this main-sequence parameterization, displaying offsets spanning $1.1$ to $30.6$ times (median: 4 times) the typical SFR at fixed stellar mass. This average offset above the main sequence is likely due to our requirement of detections of faint auroral \oiii4363\AA~and/or \oii7320,7331\AA~lines. Nebular lines are generally brighter with increasing SFR at fixed stellar mass, but in the presence of an FMR higher-sSFR galaxies tend to have lower metallicities which is also associated with brighter auroral lines due to higher $T_e$ values.

\section{Results}
\label{sec:results}

\subsection{Mass - Metallicity Relation}
\label{sec:MZR}
\begin{figure}
    \centering
    \includegraphics[width=\columnwidth]{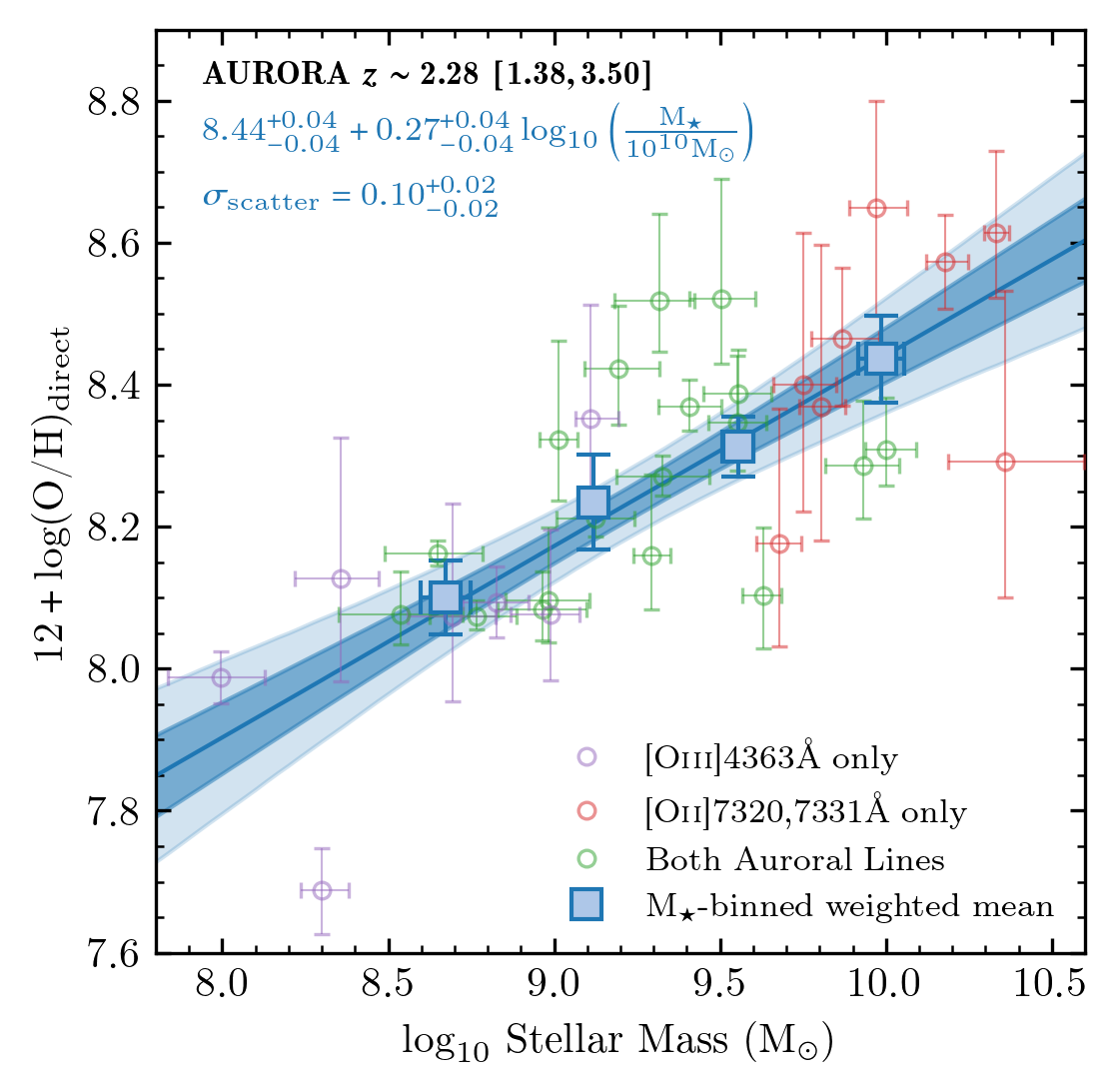}
    \caption{Mass metallicity relation for AURORA galaxies at $z \sim 2.3$ using only direct $T_e$ metallicities. \textit{Open circles} show individual galaxies color-coded based on whether \oiii4363\AA-only, \oii7320,7331\AA-only, or both oxygen auroral lines were detected. Our best-fit model (\textit{blue solid} line) is shown along with the best-fit parameters and the \textit{darker} (\textit{lighter}) shading corresponds to the $1\sigma$ ($2\sigma$) confidence regions. Weighted mean values for equally populated stellar mass bins are shown for visualization purposes, but not used in fitting.}
    \label{fig:MZR_AURORA}
\end{figure}

\begin{figure*}
    \centering
    \includegraphics[width=\textwidth]{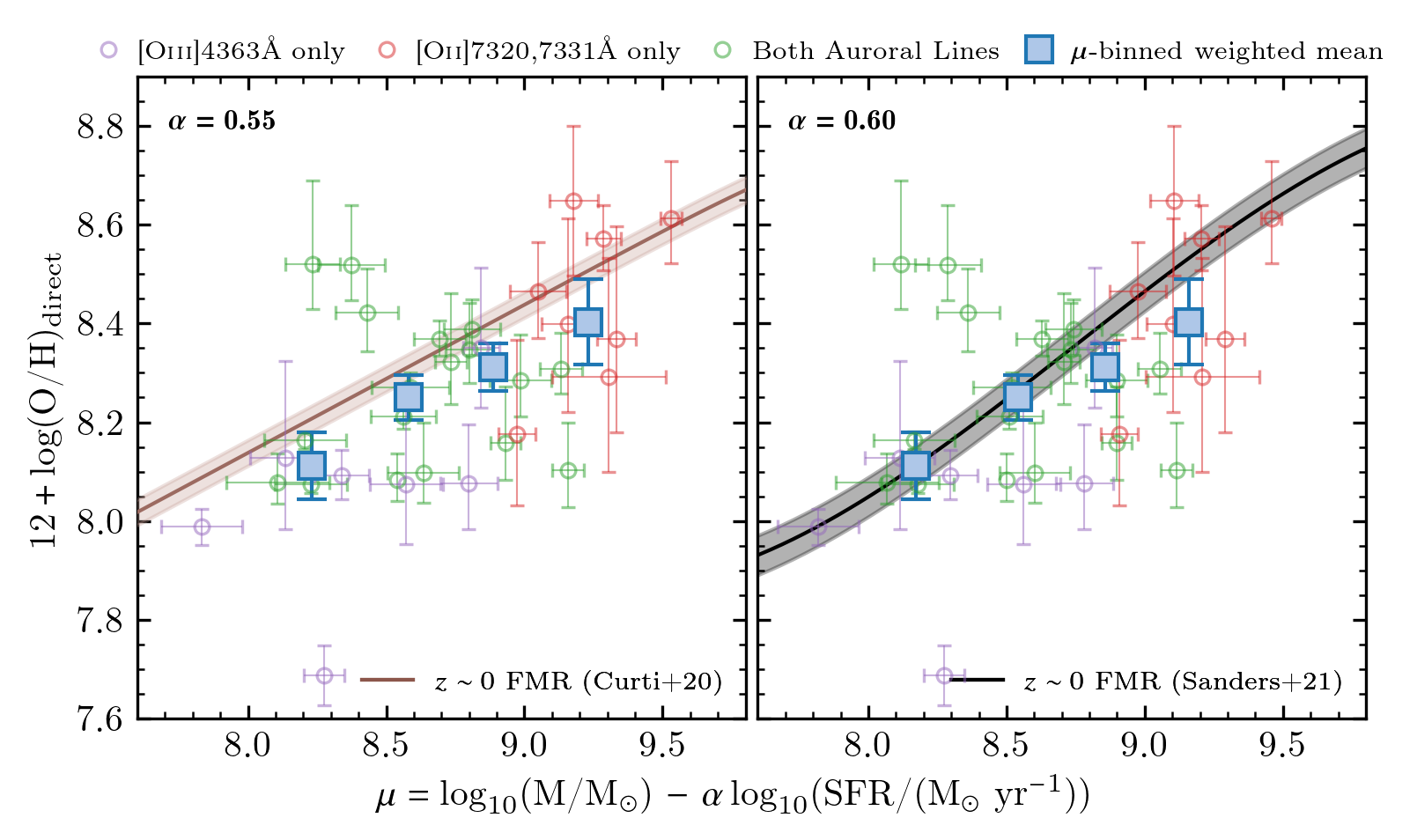}
    \caption{Fundamental Metallicity Relation with two different $\alpha$ that minimizes the scatter in \logoh~at $z \sim 0$. Our measurements show how combining both \oiii~and \oii~auroral lines results in a $\sim 1.5$ dex coverage in $\mu$ from direct $T_e$ measurements. We find our measurements are in strong agreement with the \cite{Sanders2021} FMR (\textit{right}) with only a deviation at $\mu > 8.7$ although still within $1\sigma$ errors. The \textit{left} panel shows the comparison with the \cite{Curti2020} FMR where we find agreement within $1\sigma$ but $\sim 0.07$ dex systematically lower \logoh~at fixed $\mu$~which is within the typical uncertainties of strong-line calibrations. Overall, this suggests that the same physical processes regulating star formation, stellar mass build-up, and chemical enrichment at $z \sim 0$ is in place at cosmic noon.}
    \label{fig:FMR}
\end{figure*}

Figure \ref{fig:MZR_AURORA} shows the direct-method metallicities as a function of stellar mass for our AURORA sample of 34 galaxies. The high-mass, high-metallicity regime is dominated by objects with detections of \oii$\lambda\lambda$7320,7331 only, while the lowest-metallicity sources have \oiii$\lambda$4363 only. The use of both of these oxygen auroral lines in defining the sample allows for a wider dynamic range in both \logoh~and stellar mass, significantly improving direct MZR constraints at cosmic noon.

We fit a power-law model to the 34 individual galaxies in the sample, using:
\begin{equation}
    12 + \log_{10}(\mathrm{O}/\mathrm{H}) = \gamma \log_{10}\Bigg(\frac{\mathrm{M}_\star}{10^{10}~\mathrm{M}_\odot}\Bigg) + Z_{10} + \mathcal{N}(0,\sigma_\mathrm{MZR})
\end{equation}
where $Z_{10}$ is the value of \logoh~at $10^{10}$ \msol~and $\gamma$ represents the MZR slope. The last term takes into account the intrinsic scatter of O/H at fixed $\mathrm{M}_\star$ ($\sigma_\mathrm{MZR}$). We use \texttt{emcee}, a Monte Carlo Markov Chain (MCMC) python package \citep{emcee}, to sample the posterior distribution functions of each parameter. We adopt uniform priors for each parameter bounded in ranges of $7 < Z_{10} < 9$, $0 < \gamma < 0.5$, and $0.001 < \sigma_\mathrm{MZR} < 0.5$. We consider 4000 steps with 64 walkers and burn-in $\sim 5\times$ the maximum auto-correlation time ($\approx37$~steps).  The resulting best-fit parameters are taken to be the 50th percentile for each parameter, shown in Figure \ref{fig:MZR_AURORA}. For visualization purposes, we also display inverse-variance weighted means in equally-populated bins of stellar mass. The best-fit model has a slope of $\gamma = 0.27\pm0.04$ and a normalization of \logoh$ = 8.44\pm 0.04$ at $10^{10}$ \msol, with an intrinsic scatter of $0.10\pm0.02$ dex.

\begin{figure*}
    \centering
    \includegraphics[width=\textwidth]{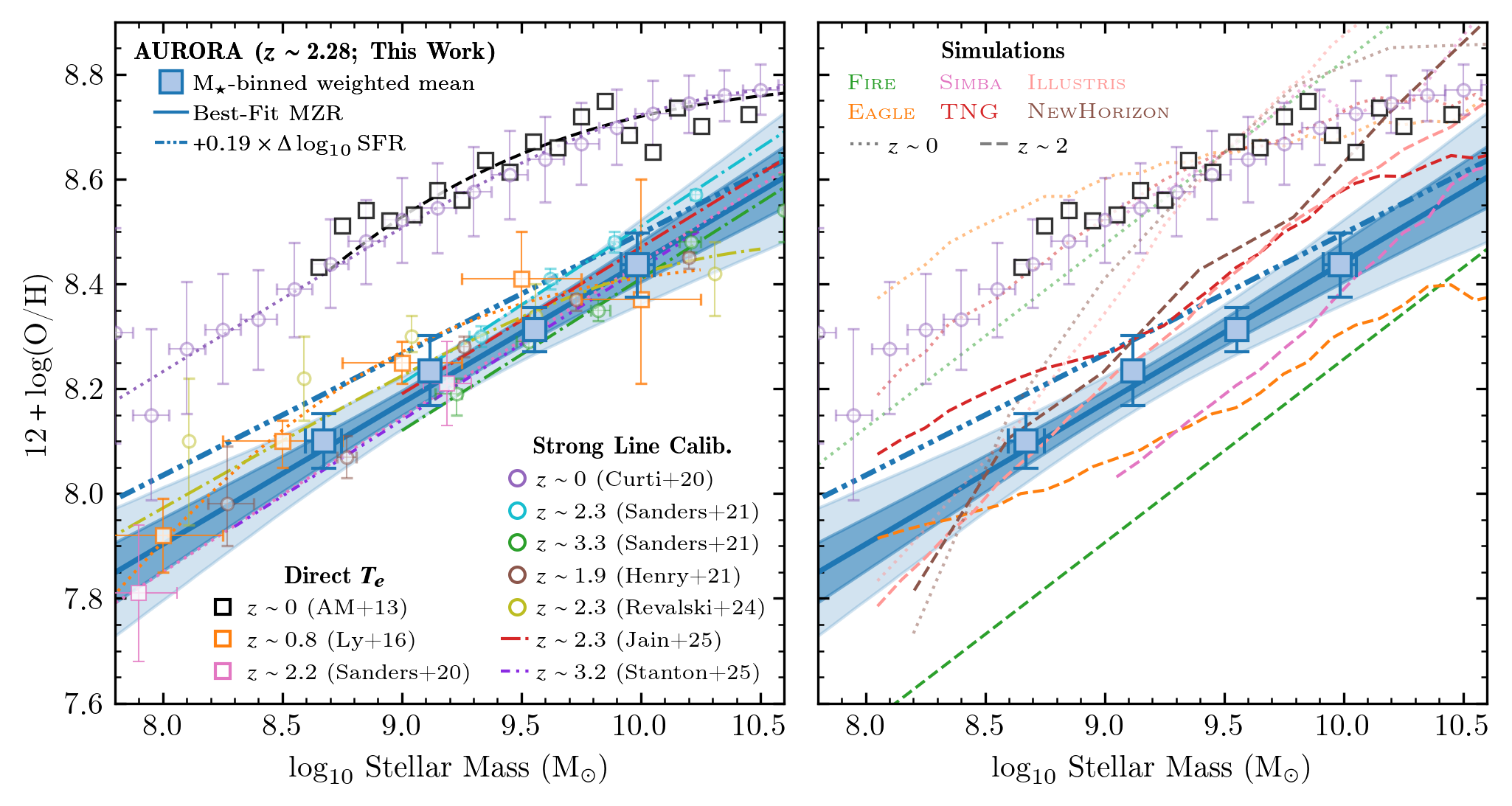}
    \caption{\textit{Left:} Our best-fit Mass -- Metallicity Relationship compared to past direct $T_e$ \citep{Curti2020,Ly2016,Sanders2020} and strong-line calibration \citep{Henry2021,Sanders2021,Revalski2024,Jain2025,Stanton2025} MZR measurements. We find our MZR is in strong agreement with past MZR studies. We also include our $0.19 \times \Delta\log_{10}\mathrm{SFR}$ corrected MZR that takes into account the bias towards high SFR, low mass systems. This MZR is found to result in increased \logoh~towards low-mass relative to past MZR studies. However, the level of correction needed is uncertain. \textit{Right:} Comparison of our MZR to different simulation predictions compiled by \cite{Garcia2025} including \texttt{FIRE} \citep{Ma2016} and \texttt{NewHorizon} \citep{Dubois2021}. Only \texttt{Illustris} and \texttt{NewHorizon} match in terms of normalization but have a steeper slope. All other simulations are found to have widely different normalizations but slopes consistent with our MZR measurement.}
    \label{fig:MZR_lit}
\end{figure*}

\subsection{Fundamental Metallicity Relation}

The Fundamental Metallicity Relation represents the three-dimensional relation among SFR, stellar mass, and gas-phase metallicity. A simple representation of the FMR is a projection onto a two-dimensional \logoh~-- $\mu$ plane defined by $\mu = \log_{10}M_* - \alpha \log_{10}\mathrm{SFR}$, where the value of $\alpha$ minimizes the scatter in \logoh~at fixed $\mu$ and is thus dependent on how metallicities were derived in different studies \citep[e.g.,][]{Mannucci2010,AM2013,Curti2020}. Our sample size with direct $T_e$ metallicities is not statistically large enough to constrain $\alpha$ directly at $z\sim2$. We instead investigate whether the $z\sim2$ population is consistent with the FMR calibrated at $z\sim0$ on average, comparing to the $z\sim0$ parameterizations of \citet{Sanders2021} with $\alpha=0.60$ and \citet{Curti2020} with $\alpha=0.55$. The results of this comparison are shown in Figure \ref{fig:FMR}. For the $z\sim2$ AURORA sample, we find median offsets in O$/$H at fixed $\mu$ of $-0.058^{+0.106}_{-0.083}$~dex relative to the \citet{Sanders2021} FMR and $-0.070^{+0.107}_{-0.085}$~dex relative to that of \citet{Curti2020}, factoring in the uncertainties of each individual \logoh\ and $\mu$ value. Our sample selection considering both the auroral \oiii4363\AA\ and \oii7320,7331\AA\ lines again leads to a wider dynamic range in properties, extending the range of $\mu$ by $\sim0.5$~dex compared to a sample selected based on \oiii4363\AA~detection alone. The binned weighted means show a correlation between \logoh~and $\mu$ with a similar slope to what is seen at $z\sim0$.

\section{Discussion}
\label{sec:discussion}

\subsection{Comparison to Observations}

We compare our best-fit direct $T_e$ $z\sim2$ MZR based purely on direct $T_e$ metallicities to observational constraints from past studies at similar redshift in the left panel of Figure \ref{fig:MZR_lit}. We find that our direct MZR measurement is consistent within $\sim1\sigma$ with past measurements at $z\sim2$ that used either direct $T_e$ \citep{Sanders2020} or strong-line calibrations \citep{Henry2021,Sanders2021,Revalski2024,Jain2025,Stanton2025}. The \citet{Sanders2020} direct $T_e$ MZR at $z\sim2$ is based on direct metallicities of 18 galaxies at $z=1.4-3.6$, derived from 7 \oiii4363\AA\ and 11 O{\sc iii}]1663\AA\ detections from ground-based spectroscopy. Our new $T_e$-based MZR represents a significant improvement over this past work, nearly doubling the sample size, achieving a much higher average S/N from space-based JWST observations, and is based entirely on rest-optical auroral lines (\oiii4363\AA~and/or \oii7320,7331\AA) to avoid the strong dust reddening systematics plaguing the use of rest-UV O{\sc iii}]1663\AA. Consequently, the uncertainties on our slope and normalization constraints are $\approx3\times$ smaller than in \citet{Sanders2020}. Other attempts to characterize the MZR via pure direct $T_e$ metallicities have been at higher redshifts and with lower precision, such as \citet{Chakraborty2025} who investigated the MZR with 42 galaxies at $z=3-10$ yielding an uncertainty on the normalization of $\approx0.2$~dex in O/H at fixed $\mathrm{M}_\star$.

However, our direct $T_e$ sample is biased toward higher than average SFR at fixed stellar mass (Figure \ref{fig:SFMS}). If a FMR exists at $z\sim2$, then a selection bias toward high SFR would introduce a corresponding bias toward low \logoh, potentially leading to an underestimate of the true MZR normalization. The magnitude of such a bias (and a corresponding correction factor) depends on the strength of the O/H--SFR anticorrelation at fixed stellar mass. \cite{Sanders2020} used the \cite{AM2013} $z\sim0$ SDSS sample to characterize this relation, finding $\Delta \log_{10} \mathrm{O/H} \propto  -0.29 \times \Delta\log_{10}\mathrm{SFR}$, where these offsets are relative to the mean MZR and star-forming main sequence, respectively. \cite{Sanders2021} reported a shallower slope of $-0.19\pm0.04$ for this relation based on a sample at $z \sim 2.3$, providing evidence that a FMR does exist at this redshift. In contrast, \cite{Korhonen2025} recently found no evidence for a significant anti-correlation between O/H and SFR at fixed mass in a $z\sim2$ sample. Observational studies are thus not in consensus regarding the strength of the secondary dependence of O/H on SFR, nor on the existence of an FMR at $z\sim2$.

Our best-fit $T_e$-based MZR shown in Figure \ref{fig:MZR_lit} may thus represent a lower limit on the actual $z\sim2$ MZR. We test how a correction to our measured O/H values assuming the \citet{Sanders2021} relation of $\Delta \log_{10} \mathrm{O/H}\propto-0.19\times\Delta\log_{10}\mathrm{SFR}$ would affect our derived $z\sim2$ MZR, adopting the \cite{Speagle2014} star-forming main sequence relation for consistency with past work. Figure \ref{fig:MZR_lit} shows this potential corrected MZR (\textit{blue dash dotted} line), for which we find a slightly shallower slope and higher normalization ($\gamma = 0.23\pm0.05$, \logoh$=8.50_{-0.04}^{+0.05}$ at 10$^{10}$ \msol). This potential correction results in a slightly shallower MZR because the lowest-mass galaxies display the largest offsets from the main sequence (Fig.~\ref{fig:SFMS}). However, this corrected MZR is $<2\sigma$ consistent with our initial MZR constraints and thus does not represent a statistically significant shift overall. Given the current disagreement in the literature about the strength (or existence) of the anticorrelation between O/H and SFR at fixed mass at $z\sim2$, further investigations are needed to understand whether our direct $T_e$ MZR is substantially biased relative to the MZR of a more representative sample.

The agreement between our $T_e$-based MZR and past strong-line-based MZRs (e.g., \citealt{Sanders2021,Jain2025}) also highlights the reasonable performance of calibrations based on local analogs of high-$z$ galaxies (e.g., \citealt{Bian2018,PerezMontero2021}). \citet{Stanton2025} uses the \cite{Scholte2025} calibration (based on \citealt{Laseter2024}), which is found to produce metallicities consistent with direct $T_e$ measurements at $z \sim 2$ \citep{Scholte2025}. \citet{Henry2021} and \citet{Revalski2024} MZRs use the \cite{Curti2017} calibration, which is based on $z\sim0$ direct $T_e$ measurements from a representative (non-analog) SDSS sample. However, the difference in metallicity when applying `normal' $z=0$ vs.\ high-redshift analog strong-line calibrations to $z\sim2$ samples is found to be $\approx0.05-0.1$~dex (see Figure 11 of \citealt{Sanders2021}). Such shifts are only at the $1-2\sigma$ level relative to the MZR normalization uncertainty (0.04~dex) with our currently small direct $T_e$ sample. We therefore conclude that the agreement between past MZR studies based on strong-line calibration and our new direct $T_e$ MZR indicates that strong-line metallicity methods perform reasonably well for cosmic noon samples.

Figure \ref{fig:FMR} shows that our $z\sim2$ sample does not display a large offset from the $z \sim 0$ FMR on average.  For the \citet{Sanders2021} FMR ($\alpha = 0.60$), we find very close agreement at $\mu < 8.7$ and an average offset of $-0.12^{+0.13}_{-0.11}$ dex in O/H at $\mu > 8.7$, though the offset in the latter range is not statistically significant. Relative to the \citet{Curti2020} $z\sim0$ FMR ($\alpha = 0.55$), the $z\sim2$ sample displays small average offset in O/H at all values of $\mu$: $-0.05^{+0.08}_{-0.06}$ dex and $-0.09^{+0.13}_{-0.11}$ dex at $\mu < 8.7$ and $\mu>8.7$, respectively. In both cases, we find that the typical offset in O/H at fixed $\mu$ is not statistically significant and suggests little FMR evolution ($\lesssim0.1$~dex in O/H at fixed stellar mass and SFR) between $z\sim0$ and $z\sim2$, in agreement with past studies at this redshift based on strong-line metallicities (e.g., \citealt{Kashino2017,Cresci2019,Sanders2021}). 

The non-evolution of the FMR and of the low-mass MZR power-law slope between $z=0$ and $z\sim2$ implies that the physical processes regulating metal enrichment, star formation activity, and stellar mass build-up at $z \sim 0$ were already in place at cosmic noon, and that $z\sim2$ galaxy growth is dominated by smooth secular processes over this mass range. Using analytical chemical evolution models applied to strong-line MZR constraints, \cite{Sanders2021} showed that the mass-scaling of outflow metal loading factors does not evolve out to $z\sim3$, and found that higher gas fractions and metal removal efficiencies at fixed mass with increasing redshift drive the decreasing MZR normalization.
Given our close agreement with the strong-line MZR and FMR of \cite{Sanders2021}, our results are consistent with the same physical picture in which feedback and metal-enriched outflows shape the MZR and regulate SFR at cosmic noon in the same manner as at $z\sim0$. The emergence of such smooth secular growth processes are also consistent with the establishment of tight cold gas scaling relations and rotationally-supported disks at $z\sim2-3$ (e.g., \citealt{Tacconi2020}; \citealt{Forster2020}, and references therein).

\subsection{Comparison to Simulations}

A direct one-to-one comparison between simulations and observations is difficult as each simulation makes distinct assumptions about stellar yields, feedback and outflow prescriptions, recycling of gas, and star-formation prescriptions, all of which can cause differences in absolute metallicity normalization. Studies based on simulations also vary in the form of metallicity that they report (e.g., weighting by mass or SFR, aperture size, stellar vs. gas-phase, star-forming vs. entire galaxy population), which may not be consistent with the metallicity probed by empirical direct $T_e$ measurements that trace ionized gas in H{\sc ii} regions. 

In order to mitigate these effects, we use the compilation of simulation results of \cite{Garcia2025} that provides MZR predictions from \texttt{EAGLE}, \texttt{SIMBA}, \texttt{Illustris}, and \texttt{IllustrisTNG}. For each simulation, \cite{Garcia2025} only selects `well-resolved' central galaxies ($>100$ star and $>500$ gas particles), limited to galaxies that have specific SFR that is $>-0.5$ dex from the median specific SFR -- stellar mass relationship in each simulation. This selection ensures that the simulated populations represent star-forming galaxies in order to make fairer comparison with observed galaxy samples used in MZR investigations. Gas-phase metallicities are defined as the mass-weighted metallicity of all star-forming gas, which should trace similar physical regions to the ionized gas in \ion{H}{2} regions. We also compare to the MZR predictions from \texttt{FIRE} \citep{Ma2016} and \texttt{NewHorizon} \citep{Dubois2021}.

To account for the different absolute metallicity scales in each simulation, we normalize the metallicities in each simulation so that their $z\sim0$ MZRs have the same metallicity at a stellar mass of $10^{9.5}$ \msol\ as the \citet{Curti2020} $z\sim0$ observational MZR. We perform the normazliation at a $10^{9.5}$ \msol~as this corresponds to the highest limiting ``well-resolved'' mass threshold among the simulation set, that of \texttt{SIMBA} \citep{Garcia2025}. This mass is also well-matched to the median mass of our $z\sim2$ $T_e$ sample. The same normalization factor is then applied to the $z \sim 2$ MZR predictions, such that the relative metallicity evolution between $z\sim0$ and $z\sim2$ is preserved.

Figure \ref{fig:MZR_lit} compares our observational direct $T_e$ MZR to the $z\sim2$ predictions from these simulations. We find that $z\sim2$ MZR normalizations present in the simulations are not consistent with our observations. \texttt{TNG}, \texttt{Illustris}, and \texttt{NewHorizon} overestimate the $z\sim2$ metallicities by $\approx0.1$~dex, while \texttt{FIRE}, \texttt{EAGLE}, and \texttt{SIMBA} fall lower than our observations by $0.1-0.2$~dex in O/H. Even at $z \sim 0$, we find that only \texttt{TNG} is in reasonable agreement with the observational \citet{Curti2020} and \citet{AM2013} MZRs at all stellar masses, while the others have either steeper or shallower MZR slopes and thus diverge at low and high masses. \texttt{FIRE}, \texttt{TNG}, and \texttt{EAGLE} have consistent $z \sim 2$ MZR slopes with our direct $T_e$ MZR, while \texttt{Illustris}, \texttt{NewHorizon}, and \texttt{SIMBA} have significantly steeper slopes.

Our results highlight that simulations have not been able to fully reproduce the observed MZR at cosmic noon and imply that changes to the feedback, gas, and/or star-formation physics in these models is needed to properly model the baryon and chemical enrichment cycles. Figure \ref{fig:MZR_lit} further shows that the disagreement between simulations and observations extends even to $z \sim 0$ where, even after our rescaling, \texttt{NewHorizon} and \texttt{Illustris} significantly diverge from the observed MZR shape. The range of different MZR evolution predictions among these 6 simulations also demonstrates that robust observational metallicity constraints across a range of redshifts can meaningfully distinguish the fidelity of different galaxy formation models. However, as discussed above, our $z \sim 2$ direct $T_e$  MZR may be biased low in \logoh~at fixed stellar mass due to a selection bias toward high SFR. The \textit{left} panel of Figure \ref{fig:MZR_lit} shows how our best-fit MZR would change if we apply a correction based on the $z\sim2$ FMR strength reported in \citet{Sanders2021}, which would lead to a $\approx 0.1$~dex increase in O/H at $10^{9.5}$ \msol~that would bring our results into better agreement with \texttt{Illustris}, \texttt{TNG}, and \texttt{NewHorizon}. This systematic shift due to potential selection bias highlights how further progress is needed on the observational side to improve the fidelity of comparisons to simulations.

\section{Conclusions}
\label{sec:conclusions}

In this Letter, we present new constraints on the mass metallicity relation and the evolution of the FMR at cosmic noon based purely on direct $T_e$ metallicities, using a sample of 34 galaxies at $z=1.4-3.5$ drawn from the AURORA Survey. Our results are as follows: 

\begin{enumerate}[label=\alph*)]
\item Our sample is made up of 34 star-forming galaxies at $1.38 \leq z \leq 3.50$ ($z_\mathrm{median} = 2.28$) with direct $T_e$ metallicities. 18 of these sources having detections of both the \oiii4363\AA~and \oii7320,7331\AA\ auroral lines. Of the remaining 16, there are 8 galaxies each with detections of \oiii4363\AA~only or \oii7320,7331\AA~only. This sample expands sample size and dynamic range in \logoh, stellar mass, and SFR used in constraining MZR and FMR relative to previous direct $T_e$-based studies at high redshift.

\item We find a best-fit MZR with slope of $\gamma=0.27^{+0.04}_{-0.04}$, and a MZR normalization of $\textrm{\logoh} = 8.44^{+0.04}_{-0.04}$ dex at $10^{10}$ \msol, with an intrinsic scatter of $0.10\pm0.02$~dex in O/H. These values are generally consistent with past $z \sim 2 - 3$ strong-line MZR measurements.

\item We compare to MZR predictions from 6 simulations and find that none of them reproduce our observed MZR normalization evolution between $z=0$ and $z\sim2$.
\texttt{Eagle}, \texttt{FIRE}, and \texttt{TNG} display $z\sim2$ MZR slopes consistent with our constraints, while \texttt{Illustris}, \texttt{SIMBA}, and \texttt{NewHorizon} have steeper slopes. These discrepancies highlight that current models do not fully reproduce the chemical enrichment and feedback processes.

\item All 34 galaxies in our sample lie on or above the star-forming main sequence relation at $z\sim2$, which may lead to an underestimate \logoh~at fixed stellar mass on average relative to a representative population of galaxies. We demonstrate that adopting the SFR-dependent correction to our measured O/H values of $+0.19\times\Delta\log_{10}\mathrm{SFR}$ correction suggested by \cite{Sanders2021} can raise our MZR by $\approx 0.1$ dex in O/H at $10^{9.5}$ \msol, bringing our MZR into better agreement with the \texttt{TNG} prediction in both slope, normalization, and evolution between $z\sim0$ and $z\sim2$. However, better constraints on the internal FMR strength (or its existence) among $z\sim2$ galaxies is required to robustly determine the degree of correction required.

\item Our direct $T_e$ measurements are consistent on average with the $z \sim 0$ FMR parameterizations of \citet{Curti2020} and \citet{Sanders2021} within 0.1~dex in O/H at fixed stellar mass and SFR. We find no statistically significant evidence for FMR evolution between $z=0$ and $z\sim2$.

\item Our findings of little-to-no FMR evolution and MZR slope evolution between $z=0$ and $z\sim2$ based purely on direct $T_e$ metallicities suggests that the mechanisms driving smooth secular chemical enrichment, star formation activity, stellar mass buildup, and metal-rich outflows were in place at cosmic noon.
\end{enumerate}

\textit{JWST} has provided the required sensitivity to robustly and efficiently investigate not only just the heavily sought after \oiii4363\AA~line, but also other key auroral lines that allow for a multi-zone analysis of electron temperatures and robust ionic abundance measurements within the interstellar medium of galaxies at cosmic noon. Future \textit{JWST} programs will expand on this work, increasing the sample sizes to improve direct $T_e$ metallicity constraints on MZR and FMR evolution and gain valuable insights into the baryon cycle during the most active period of star formation in the Universe. These measurements will also provide important constraints to further develop our chemical enrichment and feedback prescriptions in large cosmological, hydrodynamical simulations.

\section*{Acknowledgements}

This work is based on observations made with the NASA/ESA/CSA James Webb Space Telescope. The data were obtained from the Mikulski Archive for Space Telescopes at the Space Telescope Science Institute, which is operated by the Association of Universities for Research in Astronomy, Inc., under NASA contract NAS 5-03127 for JWST. These observations are associated with program JWST-GO-01914. The specific observations analyzed can be accessed via \dataset[DOI: 10.17909/hvne-7139]{https://archive.stsci.edu/doi/resolve/resolve.html?doi=10.17909/hvne-7139}.
Support for program JWST-GO-01914 was provided by NASA through a grant from the Space Telescope Science Institute, which is operated by the Association of Universities for Research in Astronomy, Inc., under NASA contract NAS 5-03127.

Some of the data products presented herein were retrieved from the Dawn JWST Archive (DJA). DJA is an initiative of the Cosmic Dawn Center (DAWN), which is funded by the Danish National Research Foundation under grant DNRF140.

This work is based on observations taken by the 3D-HST Treasury Program (GO 12177 and 12328) with the NASA/ESA HST, which is operated by the Association of Universities for Research in Astronomy, Inc., under NASA contract NAS5-26555.\\


\textit{Facilities}: JWST (NIRSpec and NIRCam), HST (ACS and WFC3)

\textit{Software}: \texttt{photutils} \citep{photutils}, \texttt{astropy} \citep{astropy:2013,astropy:2018,astropy:2022}, \texttt{numpy} \citep{numpy}
\texttt{Bagpipes} \citep{Carnall2018}, \texttt{Cloudy} \citep{Ferland2017}, \texttt{emcee} \citep{emcee}, \texttt{PyNeb} \citep{Luridiana2015}


\bibliographystyle{aasjournal}
\bibliography{AURORA_MZR} 

@ARTICLE{AM2013,
       author = {{Andrews}, Brett H. and {Martini}, Paul},
        title = "{The Mass-Metallicity Relation with the Direct Method on Stacked Spectra of SDSS Galaxies}",
      journal = {\apj},
         year = 2013,
        month = mar,
       volume = {765},
       number = {2},
          eid = {140},
        pages = {140},
          doi = {10.1088/0004-637X/765/2/140},
archivePrefix = {arXiv},
       eprint = {1211.3418},
 primaryClass = {astro-ph.CO},
       adsurl = {https://ui.adsabs.harvard.edu/abs/2013ApJ...765..140A}
}

@ARTICLE{Lodders2025,
       author = {{Lodders}, K. and {Bergemann}, M. and {Palme}, H.},
        title = "{Solar System Elemental Abundances from the Solar Photosphere and CI-Chondrites}",
      journal = {\ssr},
         year = 2025,
        month = mar,
       volume = {221},
       number = {2},
          eid = {23},
        pages = {23},
          doi = {10.1007/s11214-025-01146-w},
archivePrefix = {arXiv},
       eprint = {2502.10575},
 primaryClass = {astro-ph.SR},
       adsurl = {https://ui.adsabs.harvard.edu/abs/2025SSRv..221...23L}
}

@ARTICLE{Magg2022,
       author = {{Magg}, Ekaterina and {Bergemann}, Maria and {Serenelli}, Aldo and {Bautista}, Manuel and {Plez}, Bertrand and {Heiter}, Ulrike and {Gerber}, Jeffrey M. and {Ludwig}, Hans-G{\"u}nter and {Basu}, Sarbani and {Ferguson}, Jason W. and {Gallego}, Helena Carvajal and {Gamrath}, S{\'e}bastien and {Palmeri}, Patrick and {Quinet}, Pascal},
        title = "{Observational constraints on the origin of the elements. IV. Standard composition of the Sun}",
      journal = {\aap},
         year = 2022,
        month = may,
       volume = {661},
          eid = {A140},
        pages = {A140},
          doi = {10.1051/0004-6361/202142971},
archivePrefix = {arXiv},
       eprint = {2203.02255},
 primaryClass = {astro-ph.SR},
       adsurl = {https://ui.adsabs.harvard.edu/abs/2022A&A...661A.140M}
}

@ARTICLE{Asplund2021,
       author = {{Asplund}, M. and {Amarsi}, A.~M. and {Grevesse}, N.},
        title = "{The chemical make-up of the Sun: A 2020 vision}",
      journal = {\aap},
         year = 2021,
        month = sep,
       volume = {653},
          eid = {A141},
        pages = {A141},
          doi = {10.1051/0004-6361/202140445},
archivePrefix = {arXiv},
       eprint = {2105.01661},
 primaryClass = {astro-ph.SR},
       adsurl = {https://ui.adsabs.harvard.edu/abs/2021A&A...653A.141A}
}

@ARTICLE{Barro2019,
       author = {{Barro}, Guillermo and {P{\'e}rez-Gonz{\'a}lez}, Pablo G. and {Cava}, Antonio and {Brammer}, Gabriel and {Pandya}, Viraj and {Eliche Moral}, Carmen and {Esquej}, Pilar and {Dom{\'\i}nguez-S{\'a}nchez}, Helena and {Alcalde Pampliega}, Belen and {Guo}, Yicheng and {Koekemoer}, Anton M. and {Trump}, Jonathan R. and {Ashby}, Matthew L.~N. and {Cardiel}, Nicolas and {Castellano}, Marco and {Conselice}, Christopher J. and {Dickinson}, Mark E. and {Dolch}, Timothy and {Donley}, Jennifer L. and {Espino Briones}, N{\'e}stor and {Faber}, Sandra M. and {Fazio}, Giovanni G. and {Ferguson}, Henry and {Finkelstein}, Steve and {Fontana}, Adriano and {Galametz}, Audrey and {Gardner}, Jonathan P. and {Gawiser}, Eric and {Giavalisco}, Mauro and {Grazian}, Andrea and {Grogin}, Norman A. and {Hathi}, Nimish P. and {Hemmati}, Shoubaneh and {Hern{\'a}n-Caballero}, Antonio and {Kocevski}, Dale and {Koo}, David C. and {Kodra}, Dritan and {Lee}, Kyoung-Soo and {Lin}, Lihwai and {Lucas}, Ray A. and {Mobasher}, Bahram and {McGrath}, Elizabeth J. and {Nandra}, Kirpal and {Nayyeri}, Hooshang and {Newman}, Jeffrey A. and {Pforr}, Janine and {Peth}, Michael and {Rafelski}, Marc and {Rodr{\'\i}guez-Munoz}, Lucia and {Salvato}, Mara and {Stefanon}, Mauro and {van der Wel}, Arjen and {Willner}, Steven P. and {Wiklind}, Tommy and {Wuyts}, Stijn},
        title = "{The CANDELS/SHARDS Multiwavelength Catalog in GOODS-N: Photometry, Photometric Redshifts, Stellar Masses, Emission-line Fluxes, and Star Formation Rates}",
      journal = {\apjs},
         year = 2019,
        month = aug,
       volume = {243},
       number = {2},
          eid = {22},
        pages = {22},
          doi = {10.3847/1538-4365/ab23f2},
archivePrefix = {arXiv},
       eprint = {1908.00569},
 primaryClass = {astro-ph.GA},
       adsurl = {https://ui.adsabs.harvard.edu/abs/2019ApJS..243...22B}
}

@ARTICLE{PerezMontero2021,
       author = {{P{\'e}rez-Montero}, E. and {Amor{\'\i}n}, R. and {S{\'a}nchez Almeida}, J. and {V{\'\i}lchez}, J.~M. and {Garc{\'\i}a-Benito}, R. and {Kehrig}, C.},
        title = "{Extreme emission-line galaxies in SDSS - I. Empirical and model-based calibrations of chemical abundances}",
      journal = {\mnras},
         year = 2021,
        month = jun,
       volume = {504},
       number = {1},
        pages = {1237-1252},
          doi = {10.1093/mnras/stab862},
archivePrefix = {arXiv},
       eprint = {2103.10464},
 primaryClass = {astro-ph.GA},
       adsurl = {https://ui.adsabs.harvard.edu/abs/2021MNRAS.504.1237P}
}

@ARTICLE{Bian2018,
       author = {{Bian}, Fuyan and {Kewley}, Lisa J. and {Dopita}, Michael A.},
        title = "{{\textquotedblleft}Direct{\textquotedblright} Gas-phase Metallicity in Local Analogs of High-redshift Galaxies: Empirical Metallicity Calibrations for High-redshift Star-forming Galaxies}",
      journal = {\apj},
         year = 2018,
        month = jun,
       volume = {859},
       number = {2},
          eid = {175},
        pages = {175},
          doi = {10.3847/1538-4357/aabd74},
archivePrefix = {arXiv},
       eprint = {1805.08224},
 primaryClass = {astro-ph.GA},
       adsurl = {https://ui.adsabs.harvard.edu/abs/2018ApJ...859..175B}
}

@software{Brammer2023,
       author = {{Brammer}, Gabriel},
        title = "{grizli}",
         year = 2023,
        month = sep,
          eid = {10.5281/zenodo.8370018},
          doi = {10.5281/zenodo.8370018},
      version = {1.9.11},
    publisher = {Zenodo},
       adsurl = {https://ui.adsabs.harvard.edu/abs/2023zndo...8370018B}
}

@ARTICLE{Calzetti2000,
       author = {{Calzetti}, Daniela and {Armus}, Lee and {Bohlin}, Ralph C. and {Kinney}, Anne L. and {Koornneef}, Jan and {Storchi-Bergmann}, Thaisa},
        title = "{The Dust Content and Opacity of Actively Star-forming Galaxies}",
      journal = {\apj},
         year = 2000,
        month = apr,
       volume = {533},
       number = {2},
        pages = {682-695},
          doi = {10.1086/308692},
archivePrefix = {arXiv},
       eprint = {astro-ph/9911459},
 primaryClass = {astro-ph},
       adsurl = {https://ui.adsabs.harvard.edu/abs/2000ApJ...533..682C}
}

@ARTICLE{Campbell1986,
       author = {{Campbell}, Alison and {Terlevich}, Roberto and {Melnick}, Jorge},
        title = "{The stellar populations and evolution of H II galaxies - I. High signal-to-noise optical spectroscopy.}",
      journal = {\mnras},
         year = 1986,
        month = dec,
       volume = {223},
        pages = {811-825},
          doi = {10.1093/mnras/223.4.811},
       adsurl = {https://ui.adsabs.harvard.edu/abs/1986MNRAS.223..811C}
}

@ARTICLE{Cardelli1989,
       author = {{Cardelli}, Jason A. and {Clayton}, Geoffrey C. and {Mathis}, John S.},
        title = "{The Relationship between Infrared, Optical, and Ultraviolet Extinction}",
      journal = {\apj},
         year = 1989,
        month = oct,
       volume = {345},
        pages = {245},
          doi = {10.1086/167900},
       adsurl = {https://ui.adsabs.harvard.edu/abs/1989ApJ...345..245C}
}

@ARTICLE{Carnall2018,
       author = {{Carnall}, A.~C. and {McLure}, R.~J. and {Dunlop}, J.~S. and {Dav{\'e}}, R.},
        title = "{Inferring the star formation histories of massive quiescent galaxies with BAGPIPES: evidence for multiple quenching mechanisms}",
      journal = {\mnras},
         year = 2018,
        month = nov,
       volume = {480},
       number = {4},
        pages = {4379-4401},
          doi = {10.1093/mnras/sty2169},
archivePrefix = {arXiv},
       eprint = {1712.04452},
 primaryClass = {astro-ph.GA},
       adsurl = {https://ui.adsabs.harvard.edu/abs/2018MNRAS.480.4379C}
}

@ARTICLE{Cataldi2025,
       author = {{Cataldi}, E. and {Belfiore}, F. and {Curti}, M. and {Moreschini}, B. and {Mannucci}, F. and {D'Amato}, Q. and {Cresci}, G. and {Feltre}, A. and {Ginolfi}, M. and {Marconi}, A. and {Amiri}, A. and {Arnaboldi}, M. and {Bertola}, E. and {Bracci}, C. and {Carniani}, S. and {Ceci}, M. and {Chakraborty}, A. and {Cirasuolo}, M. and {Cullen}, F. and {Kobayashi}, C. and {Kumari}, N. and {Maiolino}, R. and {Marconcini}, C. and {Scialpi}, M. and {Ulivi}, L.},
        title = "{MARTA: Temperature-temperature relationships and strong-line metallicity calibrations from multiple auroral-line detections at cosmic noon}",
      journal = {arXiv e-prints},
         year = 2025,
        month = apr,
          eid = {arXiv:2504.03839},
        pages = {arXiv:2504.03839},
          doi = {10.48550/arXiv.2504.03839},
archivePrefix = {arXiv},
       eprint = {2504.03839},
 primaryClass = {astro-ph.GA},
       adsurl = {https://ui.adsabs.harvard.edu/abs/2025arXiv250403839C}
}

@ARTICLE{Chakraborty2025,
       author = {{Chakraborty}, Priyanka and {Sarkar}, Arnab and {Smith}, Randall and {Ferland}, Gary J. and {McDonald}, Michael and {Forman}, William and {Vogelsberger}, Mark and {Torrey}, Paul and {Garcia}, Alex M. and {Bautz}, Mark and {Foster}, Adam and {Miller}, Eric and {Grant}, Catherine},
        title = "{Unveiling the Cosmic Chemistry. II. ``Direct'' T$_{e}$-based Metallicity of Galaxies at 3 < z < 10 with JWST/NIRSpec}",
      journal = {\apj},
         year = 2025,
        month = may,
       volume = {985},
       number = {1},
          eid = {24},
        pages = {24},
          doi = {10.3847/1538-4357/adc7b5},
archivePrefix = {arXiv},
       eprint = {2412.15435},
 primaryClass = {astro-ph.GA},
       adsurl = {https://ui.adsabs.harvard.edu/abs/2025ApJ...985...24C}
}

@ARTICLE{Christensen2012,
       author = {{Christensen}, Lise and {Laursen}, Peter and {Richard}, Johan and {Hjorth}, Jens and {Milvang-Jensen}, Bo and {Dessauges-Zavadsky}, Miroslava and {Limousin}, Marceau and {Grillo}, Claudio and {Ebeling}, Harald},
        title = "{Gravitationally lensed galaxies at 2 < z < 3.5: direct abundance measurements of Ly {\ensuremath{\alpha}} emitters}",
      journal = {\mnras},
         year = 2012,
        month = dec,
       volume = {427},
       number = {3},
        pages = {1973-1982},
          doi = {10.1111/j.1365-2966.2012.22007.x},
archivePrefix = {arXiv},
       eprint = {1209.0775},
 primaryClass = {astro-ph.CO},
       adsurl = {https://ui.adsabs.harvard.edu/abs/2012MNRAS.427.1973C}
}

@ARTICLE{Clarke2024,
       author = {{Clarke}, Leonardo and {Shapley}, Alice E. and {Sanders}, Ryan L. and {Topping}, Michael W. and {Brammer}, Gabriel B. and {Bento}, Trinity and {Reddy}, Naveen A. and {Kehoe}, Emily},
        title = "{The Star-forming Main Sequence in JADES and CEERS at z > 1.4: Investigating the Burstiness of Star Formation}",
      journal = {\apj},
         year = 2024,
        month = dec,
       volume = {977},
       number = {1},
          eid = {133},
        pages = {133},
          doi = {10.3847/1538-4357/ad8ba4},
archivePrefix = {arXiv},
       eprint = {2406.05178},
 primaryClass = {astro-ph.GA},
       adsurl = {https://ui.adsabs.harvard.edu/abs/2024ApJ...977..133C}
}

@ARTICLE{Cresci2019,
       author = {{Cresci}, G. and {Mannucci}, F. and {Curti}, M.},
        title = "{Fundamental metallicity relation in CALIFA, SDSS-IV MaNGA, and high-z galaxies}",
      journal = {\aap},
         year = 2019,
        month = jul,
       volume = {627},
          eid = {A42},
        pages = {A42},
          doi = {10.1051/0004-6361/201834637},
archivePrefix = {arXiv},
       eprint = {1811.06015},
 primaryClass = {astro-ph.GA},
       adsurl = {https://ui.adsabs.harvard.edu/abs/2019A&A...627A..42C}
}

@ARTICLE{Cullen2014,
       author = {{Cullen}, F. and {Cirasuolo}, M. and {McLure}, R.~J. and {Dunlop}, J.~S. and {Bowler}, R.~A.~A.},
        title = "{The mass-metallicity-star formation rate relation at z {\ensuremath{\gtrsim}} 2 with 3D Hubble Space Telescope}",
      journal = {\mnras},
         year = 2014,
        month = may,
       volume = {440},
       number = {3},
        pages = {2300-2312},
          doi = {10.1093/mnras/stu443},
archivePrefix = {arXiv},
       eprint = {1310.0816},
 primaryClass = {astro-ph.CO},
       adsurl = {https://ui.adsabs.harvard.edu/abs/2014MNRAS.440.2300C}
}

@ARTICLE{Curti2017,
       author = {{Curti}, M. and {Cresci}, G. and {Mannucci}, F. and {Marconi}, A. and {Maiolino}, R. and {Esposito}, S.},
        title = "{New fully empirical calibrations of strong-line metallicity indicators in star-forming galaxies}",
      journal = {\mnras},
         year = 2017,
        month = feb,
       volume = {465},
       number = {2},
        pages = {1384-1400},
          doi = {10.1093/mnras/stw2766},
archivePrefix = {arXiv},
       eprint = {1610.06939},
 primaryClass = {astro-ph.GA},
       adsurl = {https://ui.adsabs.harvard.edu/abs/2017MNRAS.465.1384C}
}

@ARTICLE{Curti2020,
       author = {{Curti}, Mirko and {Mannucci}, Filippo and {Cresci}, Giovanni and {Maiolino}, Roberto},
        title = "{The mass-metallicity and the fundamental metallicity relation revisited on a fully T$_{e}$-based abundance scale for galaxies}",
      journal = {\mnras},
         year = 2020,
        month = jan,
       volume = {491},
       number = {1},
        pages = {944-964},
          doi = {10.1093/mnras/stz2910},
archivePrefix = {arXiv},
       eprint = {1910.00597},
 primaryClass = {astro-ph.GA},
       adsurl = {https://ui.adsabs.harvard.edu/abs/2020MNRAS.491..944C}
}

@ARTICLE{Curti2023,
       author = {{Curti}, Mirko and {D'Eugenio}, Francesco and {Carniani}, Stefano and {Maiolino}, Roberto and {Sandles}, Lester and {Witstok}, Joris and {Baker}, William M. and {Bennett}, Jake S. and {Piotrowska}, Joanna M. and {Tacchella}, Sandro and {Charlot}, Stephane and {Nakajima}, Kimihiko and {Maheson}, Gabriel and {Mannucci}, Filippo and {Amiri}, Amirnezam and {Arribas}, Santiago and {Belfiore}, Francesco and {Bonaventura}, Nina R. and {Bunker}, Andrew J. and {Chevallard}, Jacopo and {Cresci}, Giovanni and {Curtis-Lake}, Emma and {Hayden-Pawson}, Connor and {Jones}, Gareth C. and {Kumari}, Nimisha and {Laseter}, Isaac and {Looser}, Tobias J. and {Marconi}, Alessandro and {Maseda}, Michael V. and {Scholtz}, Jan and {Smit}, Renske and {{\"U}bler}, Hannah and {Wallace}, Imaan E.~B.},
        title = "{The chemical enrichment in the early Universe as probed by JWST via direct metallicity measurements at z {\ensuremath{\sim}} 8}",
      journal = {\mnras},
         year = 2023,
        month = jan,
       volume = {518},
       number = {1},
        pages = {425-438},
          doi = {10.1093/mnras/stac2737},
archivePrefix = {arXiv},
       eprint = {2207.12375},
 primaryClass = {astro-ph.GA},
       adsurl = {https://ui.adsabs.harvard.edu/abs/2023MNRAS.518..425C}
}

@ARTICLE{Donnan2024,
       author = {{Donnan}, C.~T. and {McLure}, R.~J. and {Dunlop}, J.~S. and {McLeod}, D.~J. and {Magee}, D. and {Arellano-C{\'o}rdova}, K.~Z. and {Barrufet}, L. and {Begley}, R. and {Bowler}, R.~A.~A. and {Carnall}, A.~C. and {Cullen}, F. and {Ellis}, R.~S. and {Fontana}, A. and {Illingworth}, G.~D. and {Grogin}, N.~A. and {Hamadouche}, M.~L. and {Koekemoer}, A.~M. and {Liu}, F.-Y. and {Mason}, C. and {Santini}, P. and {Stanton}, T.~M.},
        title = "{JWST PRIMER: a new multifield determination of the evolving galaxy UV luminosity function at redshifts z ≃ 9 - 15}",
      journal = {\mnras},
         year = 2024,
        month = sep,
       volume = {533},
       number = {3},
        pages = {3222-3237},
          doi = {10.1093/mnras/stae2037},
archivePrefix = {arXiv},
       eprint = {2403.03171},
 primaryClass = {astro-ph.GA},
       adsurl = {https://ui.adsabs.harvard.edu/abs/2024MNRAS.533.3222D}
}

@ARTICLE{Dubois2021,
       author = {{Dubois}, Yohan and {Beckmann}, Ricarda and {Bournaud}, Fr{\'e}d{\'e}ric and {Choi}, Hoseung and {Devriendt}, Julien and {Jackson}, Ryan and {Kaviraj}, Sugata and {Kimm}, Taysun and {Kraljic}, Katarina and {Laigle}, Clotilde and {Martin}, Garreth and {Park}, Min-Jung and {Peirani}, S{\'e}bastien and {Pichon}, Christophe and {Volonteri}, Marta and {Yi}, Sukyoung K.},
        title = "{Introducing the NEWHORIZON simulation: Galaxy properties with resolved internal dynamics across cosmic time}",
      journal = {\aap},
         year = 2021,
        month = jul,
       volume = {651},
          eid = {A109},
        pages = {A109},
          doi = {10.1051/0004-6361/202039429},
archivePrefix = {arXiv},
       eprint = {2009.10578},
 primaryClass = {astro-ph.GA},
       adsurl = {https://ui.adsabs.harvard.edu/abs/2021A&A...651A.109D}
}

@ARTICLE{Eisenstein2023,
       author = {{Eisenstein}, Daniel J. and {Willott}, Chris and {Alberts}, Stacey and {Arribas}, Santiago and {Bonaventura}, Nina and {Bunker}, Andrew J. and {Cameron}, Alex J. and {Carniani}, Stefano and {Charlot}, Stephane and {Curtis-Lake}, Emma and {D'Eugenio}, Francesco and {Endsley}, Ryan and {Ferruit}, Pierre and {Giardino}, Giovanna and {Hainline}, Kevin and {Hausen}, Ryan and {Jakobsen}, Peter and {Johnson}, Benjamin D. and {Maiolino}, Roberto and {Rieke}, Marcia and {Rieke}, George and {Rix}, Hans-Walter and {Robertson}, Brant and {Stark}, Daniel P. and {Tacchella}, Sandro and {Williams}, Christina C. and {Willmer}, Christopher N.~A. and {Baker}, William M. and {Baum}, Stefi and {Bhatawdekar}, Rachana and {Boyett}, Kristan and {Chen}, Zuyi and {Chevallard}, Jacopo and {Circosta}, Chiara and {Curti}, Mirko and {Danhaive}, A. Lola and {DeCoursey}, Christa and {de Graaff}, Anna and {Dressler}, Alan and {Egami}, Eiichi and {Helton}, Jakob M. and {Hviding}, Raphael E. and {Ji}, Zhiyuan and {Jones}, Gareth C. and {Kumari}, Nimisha and {L{\"u}tzgendorf}, Nora and {Laseter}, Isaac and {Looser}, Tobias J. and {Lyu}, Jianwei and {Maseda}, Michael V. and {Nelson}, Erica and {Parlanti}, Eleonora and {Perna}, Michele and {Pusk{\'a}s}, D{\'a}vid and {Rawle}, Tim and {Rodr{\'\i}guez Del Pino}, Bruno and {Sandles}, Lester and {Saxena}, Aayush and {Scholtz}, Jan and {Sharpe}, Katherine and {Shivaei}, Irene and {Silcock}, Maddie S. and {Simmonds}, Charlotte and {Skarbinski}, Maya and {Smit}, Renske and {Stone}, Meredith and {Suess}, Katherine A. and {Sun}, Fengwu and {Tang}, Mengtao and {Topping}, Michael W. and {{\"U}bler}, Hannah and {Villanueva}, Natalia C. and {Wallace}, Imaan E.~B. and {Whitler}, Lily and {Witstok}, Joris and {Woodrum}, Charity},
        title = "{Overview of the JWST Advanced Deep Extragalactic Survey (JADES)}",
      journal = {arXiv e-prints},
         year = 2023,
        month = jun,
          eid = {arXiv:2306.02465},
        pages = {arXiv:2306.02465},
          doi = {10.48550/arXiv.2306.02465},
archivePrefix = {arXiv},
       eprint = {2306.02465},
 primaryClass = {astro-ph.GA},
       adsurl = {https://ui.adsabs.harvard.edu/abs/2023arXiv230602465E}
}

@ARTICLE{Erb2006,
       author = {{Erb}, Dawn K. and {Shapley}, Alice E. and {Pettini}, Max and {Steidel}, Charles C. and {Reddy}, Naveen A. and {Adelberger}, Kurt L.},
        title = "{The Mass-Metallicity Relation at z>\raisebox{-0.5ex}\textasciitilde2}",
      journal = {\apj},
         year = 2006,
        month = jun,
       volume = {644},
       number = {2},
        pages = {813-828},
          doi = {10.1086/503623},
archivePrefix = {arXiv},
       eprint = {astro-ph/0602473},
 primaryClass = {astro-ph},
       adsurl = {https://ui.adsabs.harvard.edu/abs/2006ApJ...644..813E}
}

@ARTICLE{Ferland2017,
       author = {{Ferland}, G.~J. and {Chatzikos}, M. and {Guzm{\'a}n}, F. and {Lykins}, M.~L. and {van Hoof}, P.~A.~M. and {Williams}, R.~J.~R. and {Abel}, N.~P. and {Badnell}, N.~R. and {Keenan}, F.~P. and {Porter}, R.~L. and {Stancil}, P.~C.},
        title = "{The 2017 Release Cloudy}",
      journal = {\rmxaa},
         year = 2017,
        month = oct,
       volume = {53},
        pages = {385-438},
          doi = {10.48550/arXiv.1705.10877},
archivePrefix = {arXiv},
       eprint = {1705.10877},
 primaryClass = {astro-ph.GA},
       adsurl = {https://ui.adsabs.harvard.edu/abs/2017RMxAA..53..385F}
}

@ARTICLE{Forster2020,
       author = {{F{\"o}rster Schreiber}, Natascha M. and {Wuyts}, Stijn},
        title = "{Star-Forming Galaxies at Cosmic Noon}",
      journal = {\araa},
         year = 2020,
        month = aug,
       volume = {58},
        pages = {661-725},
          doi = {10.1146/annurev-astro-032620-021910},
archivePrefix = {arXiv},
       eprint = {2010.10171},
 primaryClass = {astro-ph.GA},
       adsurl = {https://ui.adsabs.harvard.edu/abs/2020ARA&A..58..661F}
}

@ARTICLE{emcee,
       author = {{Foreman-Mackey}, Daniel and {Hogg}, David W. and {Lang}, Dustin and {Goodman}, Jonathan},
        title = "{emcee: The MCMC Hammer}",
      journal = {\pasp},
         year = 2013,
        month = mar,
       volume = {125},
       number = {925},
        pages = {306},
          doi = {10.1086/670067},
archivePrefix = {arXiv},
       eprint = {1202.3665},
 primaryClass = {astro-ph.IM},
       adsurl = {https://ui.adsabs.harvard.edu/abs/2013PASP..125..306F}
}

@ARTICLE{Garcia2025,
       author = {{Garcia}, Alex M. and {Torrey}, Paul and {Ellison}, Sara L. and {Grasha}, Kathryn and {Chen}, Qian-Hui and {Hemler}, Z.~S. and {Zimmerman}, Dhruv T. and {Wright}, Ruby J. and {Zovaro}, Henry R.~M. and {Nelson}, Erica J. and {Sanders}, Ryan L. and {Kewley}, Lisa J. and {Hernquist}, Lars},
        title = "{Does the fundamental metallicity relation evolve with redshift? - II. The evolution in normalization of the mass-metallicity relation}",
      journal = {\mnras},
         year = 2025,
        month = jan,
       volume = {536},
       number = {1},
        pages = {119-144},
          doi = {10.1093/mnras/stae2587},
archivePrefix = {arXiv},
       eprint = {2407.06254},
 primaryClass = {astro-ph.GA},
       adsurl = {https://ui.adsabs.harvard.edu/abs/2025MNRAS.536..119G}
}

@ARTICLE{Gburek2019,
       author = {{Gburek}, Timothy and {Siana}, Brian and {Alavi}, Anahita and {Emami}, Najmeh and {Richard}, Johan and {Freeman}, William R. and {Stark}, Daniel P. and {Snapp-Kolas}, Christopher and {Lucero}, Breanna},
        title = "{The Detection of [O III] {\ensuremath{\lambda}}4363 in a Lensed, Dwarf Galaxy at z = 2.59: Testing Metallicity Indicators and Scaling Relations at High Redshift and Low Mass}",
      journal = {\apj},
         year = 2019,
        month = dec,
       volume = {887},
       number = {2},
          eid = {168},
        pages = {168},
          doi = {10.3847/1538-4357/ab5713},
archivePrefix = {arXiv},
       eprint = {1906.11849},
 primaryClass = {astro-ph.GA},
       adsurl = {https://ui.adsabs.harvard.edu/abs/2019ApJ...887..168G}
}

@ARTICLE{Gburek2023,
       author = {{Gburek}, Timothy and {Siana}, Brian and {Alavi}, Anahita and {Emami}, Najmeh and {Richard}, Johan and {Freeman}, William R. and {Stark}, Daniel P. and {Snapp-Kolas}, Christopher},
        title = "{The Direct-method Oxygen Abundance of Typical Dwarf Galaxies at Cosmic High Noon}",
      journal = {\apj},
         year = 2023,
        month = may,
       volume = {948},
       number = {2},
          eid = {108},
        pages = {108},
          doi = {10.3847/1538-4357/acb153},
archivePrefix = {arXiv},
       eprint = {2208.05976},
 primaryClass = {astro-ph.GA},
       adsurl = {https://ui.adsabs.harvard.edu/abs/2023ApJ...948..108G}
}

@ARTICLE{Grogin2011,
       author = {{Grogin}, Norman A. and {Kocevski}, Dale D. and {Faber}, S.~M. and {Ferguson}, Henry C. and {Koekemoer}, Anton M. and {Riess}, Adam G. and {Acquaviva}, Viviana and {Alexander}, David M. and {Almaini}, Omar and {Ashby}, Matthew L.~N. and {Barden}, Marco and {Bell}, Eric F. and {Bournaud}, Fr{\'e}d{\'e}ric and {Brown}, Thomas M. and {Caputi}, Karina I. and {Casertano}, Stefano and {Cassata}, Paolo and {Castellano}, Marco and {Challis}, Peter and {Chary}, Ranga-Ram and {Cheung}, Edmond and {Cirasuolo}, Michele and {Conselice}, Christopher J. and {Roshan Cooray}, Asantha and {Croton}, Darren J. and {Daddi}, Emanuele and {Dahlen}, Tomas and {Dav{\'e}}, Romeel and {de Mello}, Du{\'\i}lia F. and {Dekel}, Avishai and {Dickinson}, Mark and {Dolch}, Timothy and {Donley}, Jennifer L. and {Dunlop}, James S. and {Dutton}, Aaron A. and {Elbaz}, David and {Fazio}, Giovanni G. and {Filippenko}, Alexei V. and {Finkelstein}, Steven L. and {Fontana}, Adriano and {Gardner}, Jonathan P. and {Garnavich}, Peter M. and {Gawiser}, Eric and {Giavalisco}, Mauro and {Grazian}, Andrea and {Guo}, Yicheng and {Hathi}, Nimish P. and {H{\"a}ussler}, Boris and {Hopkins}, Philip F. and {Huang}, Jia-Sheng and {Huang}, Kuang-Han and {Jha}, Saurabh W. and {Kartaltepe}, Jeyhan S. and {Kirshner}, Robert P. and {Koo}, David C. and {Lai}, Kamson and {Lee}, Kyoung-Soo and {Li}, Weidong and {Lotz}, Jennifer M. and {Lucas}, Ray A. and {Madau}, Piero and {McCarthy}, Patrick J. and {McGrath}, Elizabeth J. and {McIntosh}, Daniel H. and {McLure}, Ross J. and {Mobasher}, Bahram and {Moustakas}, Leonidas A. and {Mozena}, Mark and {Nandra}, Kirpal and {Newman}, Jeffrey A. and {Niemi}, Sami-Matias and {Noeske}, Kai G. and {Papovich}, Casey J. and {Pentericci}, Laura and {Pope}, Alexandra and {Primack}, Joel R. and {Rajan}, Abhijith and {Ravindranath}, Swara and {Reddy}, Naveen A. and {Renzini}, Alvio and {Rix}, Hans-Walter and {Robaina}, Aday R. and {Rodney}, Steven A. and {Rosario}, David J. and {Rosati}, Piero and {Salimbeni}, Sara and {Scarlata}, Claudia and {Siana}, Brian and {Simard}, Luc and {Smidt}, Joseph and {Somerville}, Rachel S. and {Spinrad}, Hyron and {Straughn}, Amber N. and {Strolger}, Louis-Gregory and {Telford}, Olivia and {Teplitz}, Harry I. and {Trump}, Jonathan R. and {van der Wel}, Arjen and {Villforth}, Carolin and {Wechsler}, Risa H. and {Weiner}, Benjamin J. and {Wiklind}, Tommy and {Wild}, Vivienne and {Wilson}, Grant and {Wuyts}, Stijn and {Yan}, Hao-Jing and {Yun}, Min S.},
        title = "{CANDELS: The Cosmic Assembly Near-infrared Deep Extragalactic Legacy Survey}",
      journal = {\apjs},
         year = 2011,
        month = dec,
       volume = {197},
       number = {2},
          eid = {35},
        pages = {35},
          doi = {10.1088/0067-0049/197/2/35},
archivePrefix = {arXiv},
       eprint = {1105.3753},
 primaryClass = {astro-ph.CO},
       adsurl = {https://ui.adsabs.harvard.edu/abs/2011ApJS..197...35G}
}

@ARTICLE{Hayashi2015,
       author = {{Hayashi}, Masao and {Ly}, Chun and {Shimasaku}, Kazuhiro and {Motohara}, Kentaro and {Malkan}, Matthew A. and {Nagao}, Tohru and {Kashikawa}, Nobunari and {Goto}, Ryosuke and {Naito}, Yoshiaki},
        title = "{Physical conditions of the interstellar medium in star-forming galaxies at z {\ensuremath{\sim}} 1.5}",
      journal = {\pasj},
         year = 2015,
        month = oct,
       volume = {67},
       number = {5},
          eid = {80},
        pages = {80},
          doi = {10.1093/pasj/psv041},
archivePrefix = {arXiv},
       eprint = {1504.05589},
 primaryClass = {astro-ph.GA},
       adsurl = {https://ui.adsabs.harvard.edu/abs/2015PASJ...67...80H}
}

@ARTICLE{Henry2021,
       author = {{Henry}, Alaina and {Rafelski}, Marc and {Sunnquist}, Ben and {Pirzkal}, Norbert and {Pacifici}, Camilla and {Atek}, Hakim and {Bagley}, Micaela and {Baronchelli}, Ivano and {Barro}, Guillermo and {Bunker}, Andrew J. and {Colbert}, James and {Dai}, Y. Sophia and {Elmegreen}, Bruce G. and {Elmegreen}, Debra Meloy and {Finkelstein}, Steven and {Kocevski}, Dale and {Koekemoer}, Anton and {Malkan}, Matthew and {Martin}, Crystal L. and {Mehta}, Vihang and {Pahl}, Anthony and {Papovich}, Casey and {Rutkowski}, Michael and {S{\'a}nchez Almeida}, Jorge and {Scarlata}, Claudia and {Snyder}, Gregory and {Teplitz}, Harry},
        title = "{The Mass-Metallicity Relation at z   1-2 and Its Dependence on the Star Formation Rate}",
      journal = {\apj},
         year = 2021,
        month = oct,
       volume = {919},
       number = {2},
          eid = {143},
        pages = {143},
          doi = {10.3847/1538-4357/ac1105},
archivePrefix = {arXiv},
       eprint = {2107.00672},
 primaryClass = {astro-ph.GA},
       adsurl = {https://ui.adsabs.harvard.edu/abs/2021ApJ...919..143H}
}

@ARTICLE{Horne1986,
       author = {{Horne}, K.},
        title = "{An optimal extraction algorithm for CCD spectroscopy.}",
      journal = {\pasp},
         year = 1986,
        month = jun,
       volume = {98},
        pages = {609-617},
          doi = {10.1086/131801},
       adsurl = {https://ui.adsabs.harvard.edu/abs/1986PASP...98..609H}
}

@ARTICLE{Isobe2023,
       author = {{Isobe}, Yuki and {Ouchi}, Masami and {Nakajima}, Kimihiko and {Harikane}, Yuichi and {Ono}, Yoshiaki and {Xu}, Yi and {Zhang}, Yechi and {Umeda}, Hiroya},
        title = "{Redshift Evolution of Electron Density in the Interstellar Medium at z   0-9 Uncovered with JWST/NIRSpec Spectra and Line-spread Function Determinations}",
      journal = {\apj},
         year = 2023,
        month = oct,
       volume = {956},
       number = {2},
          eid = {139},
        pages = {139},
          doi = {10.3847/1538-4357/acf376},
archivePrefix = {arXiv},
       eprint = {2301.06811},
 primaryClass = {astro-ph.GA},
       adsurl = {https://ui.adsabs.harvard.edu/abs/2023ApJ...956..139I}
}

@ARTICLE{Jain2025,
       author = {{Jain}, Shweta and {Sanders}, Ryan L. and {Khostovan}, Ali Ahmad and {Jones}, Tucker and {Shapley}, Alice E. and {Reddy}, Naveen A. and {Garcia}, Alex M. and {Torrey}, Paul and {Coil}, Alison},
        title = "{A Uniform Analysis of Gas-phase Metallicity Evolution with 1-3 Gyr Time Sampling over the Past 12 Billion Years}",
      journal = {arXiv e-prints},
         year = 2025,
        month = aug,
          eid = {arXiv:2508.18369},
        pages = {arXiv:2508.18369},
          doi = {10.48550/arXiv.2508.18369},
archivePrefix = {arXiv},
       eprint = {2508.18369},
 primaryClass = {astro-ph.GA},
       adsurl = {https://ui.adsabs.harvard.edu/abs/2025arXiv250818369J}
}

@ARTICLE{Kashino2017,
       author = {{Kashino}, D. and {Silverman}, J.~D. and {Sanders}, D. and {Kartaltepe}, J.~S. and {Daddi}, E. and {Renzini}, A. and {Valentino}, F. and {Rodighiero}, G. and {Juneau}, S. and {Kewley}, L.~J. and {Zahid}, H.~J. and {Arimoto}, N. and {Nagao}, T. and {Chu}, J. and {Sugiyama}, N. and {Civano}, F. and {Ilbert}, O. and {Kajisawa}, M. and {Le F{\`e}vre}, O. and {Maier}, C. and {Masters}, D. and {Miyaji}, T. and {Onodera}, M. and {Puglisi}, A. and {Taniguchi}, Y.},
        title = "{The FMOS-COSMOS Survey of Star-forming Galaxies at z {\ensuremath{\approx}} 1.6. IV. Excitation State and Chemical Enrichment of the Interstellar Medium}",
      journal = {\apj},
         year = 2017,
        month = jan,
       volume = {835},
       number = {1},
          eid = {88},
        pages = {88},
          doi = {10.3847/1538-4357/835/1/88},
archivePrefix = {arXiv},
       eprint = {1604.06802},
 primaryClass = {astro-ph.GA},
       adsurl = {https://ui.adsabs.harvard.edu/abs/2017ApJ...835...88K}
}

@ARTICLE{Kewley2002,
       author = {{Kewley}, L.~J. and {Dopita}, M.~A.},
        title = "{Using Strong Lines to Estimate Abundances in Extragalactic H II Regions and Starburst Galaxies}",
      journal = {\apjs},
         year = 2002,
        month = sep,
       volume = {142},
       number = {1},
        pages = {35-52},
          doi = {10.1086/341326},
archivePrefix = {arXiv},
       eprint = {astro-ph/0206495},
 primaryClass = {astro-ph},
       adsurl = {https://ui.adsabs.harvard.edu/abs/2002ApJS..142...35K}
}

@ARTICLE{Kewley2019,
       author = {{Kewley}, Lisa J. and {Nicholls}, David C. and {Sutherland}, Ralph S.},
        title = "{Understanding Galaxy Evolution Through Emission Lines}",
      journal = {\araa},
         year = 2019,
        month = aug,
       volume = {57},
        pages = {511-570},
          doi = {10.1146/annurev-astro-081817-051832},
archivePrefix = {arXiv},
       eprint = {1910.09730},
 primaryClass = {astro-ph.GA},
       adsurl = {https://ui.adsabs.harvard.edu/abs/2019ARA&A..57..511K}
}

@ARTICLE{Khostovan2015,
       author = {{Khostovan}, A.~A. and {Sobral}, D. and {Mobasher}, B. and {Best}, P.~N. and {Smail}, I. and {Stott}, J.~P. and {Hemmati}, S. and {Nayyeri}, H.},
        title = "{Evolution of the H {\ensuremath{\beta}} + [O III] and [O II] luminosity functions and the [O II] star formation history of the Universe up to z {\ensuremath{\sim}} 5 from HiZELS}",
      journal = {\mnras},
         year = 2015,
        month = oct,
       volume = {452},
       number = {4},
        pages = {3948-3968},
          doi = {10.1093/mnras/stv1474},
archivePrefix = {arXiv},
       eprint = {1503.00004},
 primaryClass = {astro-ph.GA},
       adsurl = {https://ui.adsabs.harvard.edu/abs/2015MNRAS.452.3948K}
}

@ARTICLE{Khostovan2016,
       author = {{Khostovan}, A.~A. and {Sobral}, D. and {Mobasher}, B. and {Smail}, I. and {Darvish}, B. and {Nayyeri}, H. and {Hemmati}, S. and {Stott}, J.~P.},
        title = "{The nature of H{\ensuremath{\beta}}+[O III] and [O II] emitters to z {\ensuremath{\sim}} 5 with HiZELS: stellar mass functions and the evolution of EWs}",
      journal = {\mnras},
         year = 2016,
        month = dec,
       volume = {463},
       number = {3},
        pages = {2363-2382},
          doi = {10.1093/mnras/stw2174},
archivePrefix = {arXiv},
       eprint = {1604.02456},
 primaryClass = {astro-ph.GA},
       adsurl = {https://ui.adsabs.harvard.edu/abs/2016MNRAS.463.2363K}
}

@ARTICLE{Khostovan2021,
       author = {{Khostovan}, A.~A. and {Malhotra}, S. and {Rhoads}, J.~E. and {Harish}, S. and {Jiang}, C. and {Wang}, J. and {Wold}, I. and {Zheng}, Z.-Y. and {Barrientos}, L.~F. and {Coughlin}, A. and {Hu}, W. and {Infante}, L. and {Perez}, L.~A. and {Pharo}, J. and {Valdes}, F. and {Walker}, A.~R.},
        title = "{Correlations between H {\ensuremath{\alpha}} equivalent width and galaxy properties at z = 0.47: Physical or selection-driven?}",
      journal = {\mnras},
         year = 2021,
        month = may,
       volume = {503},
       number = {4},
        pages = {5115-5133},
          doi = {10.1093/mnras/stab778},
archivePrefix = {arXiv},
       eprint = {2103.10959},
 primaryClass = {astro-ph.GA},
       adsurl = {https://ui.adsabs.harvard.edu/abs/2021MNRAS.503.5115K}
}

@ARTICLE{Khostovan2024,
       author = {{Khostovan}, A.~A. and {Malhotra}, S. and {Rhoads}, J.~E. and {Sobral}, D. and {Harish}, S. and {Tilvi}, V. and {Coughlin}, A. and {Rezaee}, S.},
        title = "{Evolution of H {\ensuremath{\alpha}} equivalent widths from z {\ensuremath{\sim}} 0.4 - 2.2: implications for star formation and legacy surveys with Roman and Euclid}",
      journal = {\mnras},
         year = 2024,
        month = dec,
       volume = {535},
       number = {4},
        pages = {2903-2926},
          doi = {10.1093/mnras/stae2395},
archivePrefix = {arXiv},
       eprint = {2408.00080},
 primaryClass = {astro-ph.GA},
       adsurl = {https://ui.adsabs.harvard.edu/abs/2024MNRAS.535.2903K}
}

@ARTICLE{Khostovan2025,
       author = {{Khostovan}, Ali Ahmad and {Kartaltepe}, Jeyhan S. and {Brinch}, Malte and {Casey}, Caitlin and {Faisst}, Andreas and {Harish}, Santosh and {Gozaliasl}, Ghassem and {Onodera}, Masato and {Yabe}, Kiyoto},
        title = "{EELG1002: A Record-breaking [O III]+H{\ensuremath{\beta}} EW {\ensuremath{\sim}} 3700 {\r{A}} Galaxy at z {\ensuremath{\sim}} 0.8{\textemdash}Analog of Early Galaxies?}",
      journal = {\apj},
         year = 2025,
        month = nov,
       volume = {994},
       number = {1},
          eid = {34},
        pages = {34},
          doi = {10.3847/1538-4357/ae0330},
archivePrefix = {arXiv},
       eprint = {2411.10537},
 primaryClass = {astro-ph.GA},
       adsurl = {https://ui.adsabs.harvard.edu/abs/2025ApJ...994...34K}
}

@ARTICLE{Koekemoer2011,
       author = {{Koekemoer}, Anton M. and {Faber}, S.~M. and {Ferguson}, Henry C. and {Grogin}, Norman A. and {Kocevski}, Dale D. and {Koo}, David C. and {Lai}, Kamson and {Lotz}, Jennifer M. and {Lucas}, Ray A. and {McGrath}, Elizabeth J. and {Ogaz}, Sara and {Rajan}, Abhijith and {Riess}, Adam G. and {Rodney}, Steve A. and {Strolger}, Louis and {Casertano}, Stefano and {Castellano}, Marco and {Dahlen}, Tomas and {Dickinson}, Mark and {Dolch}, Timothy and {Fontana}, Adriano and {Giavalisco}, Mauro and {Grazian}, Andrea and {Guo}, Yicheng and {Hathi}, Nimish P. and {Huang}, Kuang-Han and {van der Wel}, Arjen and {Yan}, Hao-Jing and {Acquaviva}, Viviana and {Alexander}, David M. and {Almaini}, Omar and {Ashby}, Matthew L.~N. and {Barden}, Marco and {Bell}, Eric F. and {Bournaud}, Fr{\'e}d{\'e}ric and {Brown}, Thomas M. and {Caputi}, Karina I. and {Cassata}, Paolo and {Challis}, Peter J. and {Chary}, Ranga-Ram and {Cheung}, Edmond and {Cirasuolo}, Michele and {Conselice}, Christopher J. and {Roshan Cooray}, Asantha and {Croton}, Darren J. and {Daddi}, Emanuele and {Dav{\'e}}, Romeel and {de Mello}, Duilia F. and {de Ravel}, Loic and {Dekel}, Avishai and {Donley}, Jennifer L. and {Dunlop}, James S. and {Dutton}, Aaron A. and {Elbaz}, David and {Fazio}, Giovanni G. and {Filippenko}, Alexei V. and {Finkelstein}, Steven L. and {Frazer}, Chris and {Gardner}, Jonathan P. and {Garnavich}, Peter M. and {Gawiser}, Eric and {Gruetzbauch}, Ruth and {Hartley}, Will G. and {H{\"a}ussler}, Boris and {Herrington}, Jessica and {Hopkins}, Philip F. and {Huang}, Jia-Sheng and {Jha}, Saurabh W. and {Johnson}, Andrew and {Kartaltepe}, Jeyhan S. and {Khostovan}, Ali A. and {Kirshner}, Robert P. and {Lani}, Caterina and {Lee}, Kyoung-Soo and {Li}, Weidong and {Madau}, Piero and {McCarthy}, Patrick J. and {McIntosh}, Daniel H. and {McLure}, Ross J. and {McPartland}, Conor and {Mobasher}, Bahram and {Moreira}, Heidi and {Mortlock}, Alice and {Moustakas}, Leonidas A. and {Mozena}, Mark and {Nandra}, Kirpal and {Newman}, Jeffrey A. and {Nielsen}, Jennifer L. and {Niemi}, Sami and {Noeske}, Kai G. and {Papovich}, Casey J. and {Pentericci}, Laura and {Pope}, Alexandra and {Primack}, Joel R. and {Ravindranath}, Swara and {Reddy}, Naveen A. and {Renzini}, Alvio and {Rix}, Hans-Walter and {Robaina}, Aday R. and {Rosario}, David J. and {Rosati}, Piero and {Salimbeni}, Sara and {Scarlata}, Claudia and {Siana}, Brian and {Simard}, Luc and {Smidt}, Joseph and {Snyder}, Diana and {Somerville}, Rachel S. and {Spinrad}, Hyron and {Straughn}, Amber N. and {Telford}, Olivia and {Teplitz}, Harry I. and {Trump}, Jonathan R. and {Vargas}, Carlos and {Villforth}, Carolin and {Wagner}, Cory R. and {Wandro}, Pat and {Wechsler}, Risa H. and {Weiner}, Benjamin J. and {Wiklind}, Tommy and {Wild}, Vivienne and {Wilson}, Grant and {Wuyts}, Stijn and {Yun}, Min S.},
        title = "{CANDELS: The Cosmic Assembly Near-infrared Deep Extragalactic Legacy Survey{\textemdash}The Hubble Space Telescope Observations, Imaging Data Products, and Mosaics}",
      journal = {\apjs},
         year = 2011,
        month = dec,
       volume = {197},
       number = {2},
          eid = {36},
        pages = {36},
          doi = {10.1088/0067-0049/197/2/36},
archivePrefix = {arXiv},
       eprint = {1105.3754},
 primaryClass = {astro-ph.CO},
       adsurl = {https://ui.adsabs.harvard.edu/abs/2011ApJS..197...36K}
}

@ARTICLE{Korhonen2025,
       author = {{Korhonen Cuestas}, Nathalie A. and {Strom}, Allison L. and {Miller}, Tim B. and {Steidel}, Charles C. and {Trainor}, Ryan F. and {Rudie}, Gwen C. and {Nu{\~n}ez}, Evan Haze},
        title = "{Exploring the Relationship between Stellar Mass, Metallicity, and Star Formation Rate at z {\ensuremath{\sim}} 2.3 in KBSS-MOSFIRE}",
      journal = {\apj},
         year = 2025,
        month = may,
       volume = {984},
       number = {2},
          eid = {188},
        pages = {188},
          doi = {10.3847/1538-4357/adc5f7},
archivePrefix = {arXiv},
       eprint = {2503.10800},
 primaryClass = {astro-ph.GA},
       adsurl = {https://ui.adsabs.harvard.edu/abs/2025ApJ...984..188K}
}

@ARTICLE{Kroupa2001,
       author = {{Kroupa}, Pavel},
        title = "{On the variation of the initial mass function}",
      journal = {\mnras},
         year = 2001,
        month = apr,
       volume = {322},
       number = {2},
        pages = {231-246},
          doi = {10.1046/j.1365-8711.2001.04022.x},
archivePrefix = {arXiv},
       eprint = {astro-ph/0009005},
 primaryClass = {astro-ph},
       adsurl = {https://ui.adsabs.harvard.edu/abs/2001MNRAS.322..231K}
}

@ARTICLE{Laseter2024,
       author = {{Laseter}, Isaac H. and {Maseda}, Michael V. and {Curti}, Mirko and {Maiolino}, Roberto and {D'Eugenio}, Francesco and {Cameron}, Alex J. and {Looser}, Tobias J. and {Arribas}, Santiago and {Baker}, William M. and {Bhatawdekar}, Rachana and {Boyett}, Kristan and {Bunker}, Andrew J. and {Carniani}, Stefano and {Charlot}, Stephane and {Chevallard}, Jacopo and {Curtis-lake}, Emma and {Egami}, Eiichi and {Eisenstein}, Daniel J. and {Hainline}, Kevin and {Hausen}, Ryan and {Ji}, Zhiyuan and {Kumari}, Nimisha and {Perna}, Michele and {Rawle}, Tim and {Rix}, Hans-Walter and {Robertson}, Brant and {Rodr{\'\i}guez Del Pino}, Bruno and {Sandles}, Lester and {Scholtz}, Jan and {Smit}, Renske and {Tacchella}, Sandro and {{\"U}bler}, Hannah and {Williams}, Christina C. and {Willott}, Chris and {Witstok}, Joris},
        title = "{JADES: Detecting [OIII]{\ensuremath{\lambda}}4363 emitters and testing strong line calibrations in the high-z Universe with ultra-deep JWST/NIRSpec spectroscopy up to z {\ensuremath{\sim}} 9.5}",
      journal = {\aap},
         year = 2024,
        month = jan,
       volume = {681},
          eid = {A70},
        pages = {A70},
          doi = {10.1051/0004-6361/202347133},
archivePrefix = {arXiv},
       eprint = {2306.03120},
 primaryClass = {astro-ph.GA},
       adsurl = {https://ui.adsabs.harvard.edu/abs/2024A&A...681A..70L}
}

@ARTICLE{Leja2019,
       author = {{Leja}, Joel and {Carnall}, Adam C. and {Johnson}, Benjamin D. and {Conroy}, Charlie and {Speagle}, Joshua S.},
        title = "{How to Measure Galaxy Star Formation Histories. II. Nonparametric Models}",
      journal = {\apj},
         year = 2019,
        month = may,
       volume = {876},
       number = {1},
          eid = {3},
        pages = {3},
          doi = {10.3847/1538-4357/ab133c},
archivePrefix = {arXiv},
       eprint = {1811.03637},
 primaryClass = {astro-ph.GA},
       adsurl = {https://ui.adsabs.harvard.edu/abs/2019ApJ...876....3L}
}

@ARTICLE{Luridiana2015,
       author = {{Luridiana}, V. and {Morisset}, C. and {Shaw}, R.~A.},
        title = "{PyNeb: a new tool for analyzing emission lines. I. Code description and validation of results}",
      journal = {\aap},
         year = 2015,
        month = jan,
       volume = {573},
          eid = {A42},
        pages = {A42},
          doi = {10.1051/0004-6361/201323152},
archivePrefix = {arXiv},
       eprint = {1410.6662},
 primaryClass = {astro-ph.IM},
       adsurl = {https://ui.adsabs.harvard.edu/abs/2015A&A...573A..42L}
}

@ARTICLE{Ly2016,
       author = {{Ly}, Chun and {Malkan}, Matthew A. and {Rigby}, Jane R. and {Nagao}, Tohru},
        title = "{The Metal Abundances across Cosmic Time (MACT) Survey. II. Evolution of the Mass-metallicity Relation over 8 Billion Years, Using [OIII]4363AA-based Metallicities}",
      journal = {\apj},
         year = 2016,
        month = sep,
       volume = {828},
       number = {2},
          eid = {67},
        pages = {67},
          doi = {10.3847/0004-637X/828/2/67},
archivePrefix = {arXiv},
       eprint = {1602.01098},
 primaryClass = {astro-ph.GA},
       adsurl = {https://ui.adsabs.harvard.edu/abs/2016ApJ...828...67L}
}

@ARTICLE{Ma2016,
       author = {{Ma}, Xiangcheng and {Hopkins}, Philip F. and {Faucher-Gigu{\`e}re}, Claude-Andr{\'e} and {Zolman}, Nick and {Muratov}, Alexander L. and {Kere{\v{s}}}, Du{\v{s}}an and {Quataert}, Eliot},
        title = "{The origin and evolution of the galaxy mass-metallicity relation}",
      journal = {\mnras},
         year = 2016,
        month = feb,
       volume = {456},
       number = {2},
        pages = {2140-2156},
          doi = {10.1093/mnras/stv2659},
archivePrefix = {arXiv},
       eprint = {1504.02097},
 primaryClass = {astro-ph.GA},
       adsurl = {https://ui.adsabs.harvard.edu/abs/2016MNRAS.456.2140M}
}

@ARTICLE{Madau2014,
       author = {{Madau}, Piero and {Dickinson}, Mark},
        title = "{Cosmic Star-Formation History}",
      journal = {\araa},
         year = 2014,
        month = aug,
       volume = {52},
        pages = {415-486},
          doi = {10.1146/annurev-astro-081811-125615},
archivePrefix = {arXiv},
       eprint = {1403.0007},
 primaryClass = {astro-ph.CO},
       adsurl = {https://ui.adsabs.harvard.edu/abs/2014ARA&A..52..415M}
}

@ARTICLE{Maier2014,
       author = {{Maier}, C. and {Lilly}, S.~J. and {Ziegler}, B.~L. and {Contini}, T. and {P{\'e}rez Montero}, E. and {Peng}, Y. and {Balestra}, I.},
        title = "{The Mass-Metallicity and Fundamental Metallicity Relations at z > 2 Using Very Large Telescope and Subaru Near-infrared Spectroscopy of zCOSMOS Galaxies}",
      journal = {\apj},
         year = 2014,
        month = sep,
       volume = {792},
       number = {1},
          eid = {3},
        pages = {3},
          doi = {10.1088/0004-637X/792/1/3},
archivePrefix = {arXiv},
       eprint = {1406.6069},
 primaryClass = {astro-ph.GA},
       adsurl = {https://ui.adsabs.harvard.edu/abs/2014ApJ...792....3M}
}

@ARTICLE{Mannucci2010,
       author = {{Mannucci}, F. and {Cresci}, G. and {Maiolino}, R. and {Marconi}, A. and {Gnerucci}, A.},
        title = "{A fundamental relation between mass, star formation rate and metallicity in local and high-redshift galaxies}",
      journal = {\mnras},
         year = 2010,
        month = nov,
       volume = {408},
       number = {4},
        pages = {2115-2127},
          doi = {10.1111/j.1365-2966.2010.17291.x},
archivePrefix = {arXiv},
       eprint = {1005.0006},
 primaryClass = {astro-ph.CO},
       adsurl = {https://ui.adsabs.harvard.edu/abs/2010MNRAS.408.2115M}
}

@ARTICLE{Marino2013,
       author = {{Marino}, R.~A. and {Rosales-Ortega}, F.~F. and {S{\'a}nchez}, S.~F. and {Gil de Paz}, A. and {V{\'\i}lchez}, J. and {Miralles-Caballero}, D. and {Kehrig}, C. and {P{\'e}rez-Montero}, E. and {Stanishev}, V. and {Iglesias-P{\'a}ramo}, J. and {D{\'\i}az}, A.~I. and {Castillo-Morales}, A. and {Kennicutt}, R. and {L{\'o}pez-S{\'a}nchez}, A.~R. and {Galbany}, L. and {Garc{\'\i}a-Benito}, R. and {Mast}, D. and {Mendez-Abreu}, J. and {Monreal-Ibero}, A. and {Husemann}, B. and {Walcher}, C.~J. and {Garc{\'\i}a-Lorenzo}, B. and {Masegosa}, J. and {Del Olmo Orozco}, A. and {Mour{\~a}o}, A.~M. and {Ziegler}, B. and {Moll{\'a}}, M. and {Papaderos}, P. and {S{\'a}nchez-Bl{\'a}zquez}, P. and {Gonz{\'a}lez Delgado}, R.~M. and {Falc{\'o}n-Barroso}, J. and {Roth}, M.~M. and {van de Ven}, G. and {CALIFA Team}},
        title = "{The O3N2 and N2 abundance indicators revisited: improved calibrations based on CALIFA and T$_{e}$-based literature data}",
      journal = {\aap},
         year = 2013,
        month = nov,
       volume = {559},
          eid = {A114},
        pages = {A114},
          doi = {10.1051/0004-6361/201321956},
archivePrefix = {arXiv},
       eprint = {1307.5316},
 primaryClass = {astro-ph.CO},
       adsurl = {https://ui.adsabs.harvard.edu/abs/2013A&A...559A.114M}
}

\end{document}